\documentclass[reprint,nofootinbib,amsmath,amssymb,showkeys,aps]{revtex4-1}

\usepackage{amssymb}
\usepackage{array,multirow}
\usepackage{enumerate}
\usepackage{bm}
\usepackage{amsmath}
\usepackage{graphicx}
\usepackage{xcolor}

\usepackage[flushleft]{threeparttable}

\begin{document}

\title{A fast test to assess the impact of marginalization in Monte Carlo analyses, and its application to cosmology}

\author{Adri\`a G\'omez-Valent$^{1,2}$}\email{agvalent@roma2.infn.it}

\affiliation{$^1$ Dipartimento di Fisica, Università di Roma Tor Vergata, via della Ricerca Scientifica, 1, 00133, Roma, Italy}
\affiliation{$^2$ INFN, Sezione di Roma 2, Università di Roma Tor Vergata, via della Ricerca Scientifica, 1, 00133 Roma, Italy}

\begin{abstract}
Monte Carlo (MC) algorithms are commonly employed to explore high-dimensional parameter spaces constrained by data. All the statistical information obtained in the output of these analyses is contained in the Markov chains, which one needs to process and interpret. The marginalization technique allows us to digest these chains and compute the posterior distributions for the parameter subsets of interest. In particular, it lets us draw confidence regions in two-dimensional planes, and get the constraints for the individual parameters. It is very well known, though, that the marginalized results can suffer from volume effects, which can introduce a non-negligible bias into our conclusions. The impact of these effects are barely studied in the literature. In this paper we first illustrate the problem through a very clear and simple example in two dimensions, and suggest the use of the profile distributions (PDs) as a complementary tool to detect marginalization biases directly from the MC chains. We apply our method to four cosmological models: the standard $\Lambda$CDM, early dark energy, coupled dark energy and the Brans-Dicke model with a cosmological constant. We discuss the impact of the volume effects on each model and the cosmological tensions, using the full {\it Planck} 2018 likelihood, the Pantheon compilation of supernovae of Type Ia and data on baryon acoustic oscillations. Our test is very efficient and can be easily applied to any MC study. It allows us to estimate the PDs at a derisory computational cost not only for the main cosmological parameters, but also for the nuisance and derived ones, and to assess the need to perform a more in-depth analysis with the exact computation of the PDs.  
\end{abstract}

\keywords{Cosmology: observations -- Cosmology: theory -- cosmological parameters}

\maketitle


\section{Introduction}\label{sec:intro}

The use of Monte Carlo (MC) methods is unavoidable in the sampling of multivariate distributions in high-dimensional parameter spaces. The posterior distributions for the various subsets of parameters in these analyses are obtained from the Markov chains through marginalization. The latter allows us e.g. to draw contour plots in all the two-dimensional planes of interest and to obtain the posteriors of the individual parameters, from which one can infer their constraints at the desired confidence level. Marginalization is a practical tool that eases the visualization and interpretation of the results in Bayesian analyses. This explains its wide use. Nevertheless, it is of utmost importance to bear in mind that the statistical content is in general not conserved in the marginalization process. Volume effects can potentially bias our conclusions, sometimes in a non-negligible way. Therefore, they can have an impact on the interpretation of the results extracted from the Monte Carlo Markov chains, especially when non-Gaussianities are sizable. 

In this paper we first illustrate in Sec. \ref{sec:SecI} this fact explicitly with a very pedagogical and clear example in two dimensions, making use of a taylor-made probability distribution function (PDF). The reader will easily understand through this example that when the non-marginalized probability density has non-Gaussian features the marginalized posteriors derived from it do not show us how the parameters are distributed according to their ability to explain the data, but according to their integrated probability weight. Due to the volume effects there can be a big mismatch between these two distributions, and in some cases the posterior can hide points in parameter space that are able to fit very well the data. This is a problem if we use the marginalized results to quantify tensions between experiments or between theoretical models. Thus, the detection of biases induced by volume effects is a very important task that we should undertake before extracting definite conclusions from the marginalized results in MC analyses. A disagreement between the point in parameter space that maximizes the non-marginalized distribution and the point built with the values of the parameters that maximize the individual one-dimensional posteriors already hints the existence of these biases. It would be useful, though, to have a method that lets us assess more accurately their impact. We suggest the use of the profile distribution (PD), which is similar to the profile likelihood (see e.g. \cite{Trotta:2017wnx} and references therein), but it can incorporate the effect of a prior as well. If $\mathcal{P}(\theta_1,\theta_2)$ is the normalized distribution for the parameter sets $\theta_1$ and $\theta_2$, we can compute the PD for $\theta_1$, called $\tilde{\mathcal{P}}(\theta_1)$, by searching for the maximum of $\mathcal{P}$ for each $\theta_1$ along the directions of $\theta_2$, i.e.

\begin{equation}\label{eq:PD}
\tilde{\mathcal{P}}(\theta_1)=\max\limits_{{\theta_2}}\mathcal{P}(\theta_1,\theta_2)\,.
\end{equation}   
We will argue that the ratio 

\begin{equation}\label{eq:ratio}
R(\theta_1)=\frac{\tilde{\mathcal{P}}(\theta_1)}{\max\limits_{{\theta_1}}\tilde{\mathcal{P}}(\theta_1)}=\frac{\tilde{\mathcal{P}}(\theta_1)}{\max\limits_{{\theta_1,\theta_2}}\mathcal{P}(\theta_1,\theta_2)}
\end{equation}
can be interpreted as a probability weight for every point in the space of $\theta_1$ and, hence, we can employ this quantity to build the distribution of this parameter set. The main advantage of doing so, instead of using the posterior

\begin{equation}\label{eq:marg}
\mathcal{P}(\theta_1)=\int \mathcal{P}(\theta_1,\theta_2)\,d\theta_2\,,
\end{equation}
is that \eqref{eq:ratio} is not subject to volume effects, since we are not performing any integration. Its maximum is located at the very same value of $\theta_1$ that maximizes $\mathcal{P}$, and the PD \eqref{eq:PD} shows explicitly what values of $\theta_1$ lead to a better description of the data, taking also into account the information from the prior. The analogous quantities for $\theta_2$ or any subset built from $\theta_1$ and/or $\theta_2$ can be also computed, of course. If the parameter space is high-dimensional (as it is e.g. in the vast majority of cosmological studies), the computation of the one-dimensional PDs for all the parameters (the main, derived and nuisance ones) can be expensive in terms of computational time, and a preliminary MC is needed to better determine the parameter ranges of interest and estimate the location of the best-fit values. It is therefore useful to assess the impact of marginalization directly from our Markov chains, before performing a more exhaustive calculation of the PDs, which in some cases could be unnecessary. Proceeding in this way we can estimate not only the impact of marginalization effects on the main cosmological parameters, but also on the derived and nuisance ones, which can be useful in some studies.

\begin{figure*}[t!]
\begin{center}
\includegraphics[width=6.5in, height=2.5in]{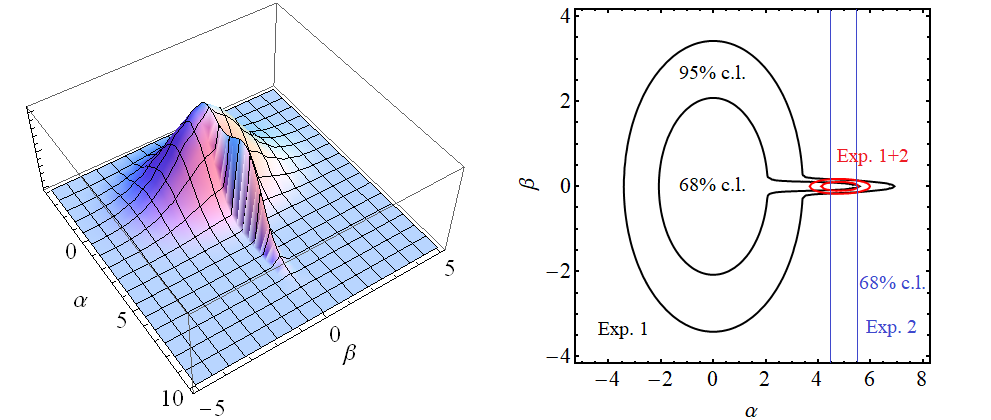}
\caption{{\it Left plot:} Shape of the probability distribution \eqref{eq:exampleDist}; {\it Right plot:} Confidence regions at $68\%$ and $95\%$ c.l. in the $(\alpha,\beta)$-plane obtained from \eqref{eq:exampleDist} (in black), together with the $68\%$-band obtained from {\it experiment 2} (in blue), and the joint constraints from the two experiments (in red). See the main text in Sec. \ref{sec:SecI} for details.}\label{fig:fig1}
\end{center}
\end{figure*} 

We do that in Sec. \ref{sec:SecII} for several cosmological models of interest, namely: the $\Lambda$CDM; early dark energy (EDE) mimicked by an ultra-light scalar field with an axion-like potential (ULA) \cite{Kamionkowski:2014zda,Poulin:2018dzj,Poulin:2018cxd}; coupled dark energy (CDE). We consider a scalar field with a Peebles-Ratra potential \cite{Peebles:1987ek,Ratra:1987rm} interacting with dark matter through a constant coupling \cite{Pettorino:2012ts,Pettorino:2013oxa,Planck:2015bue,Gomez-Valent:2020mqn}. CDE was originally proposed in \cite{Wetterich:1994bg,Amendola:1999er}; and the Brans-Dicke (BD) model \cite{BransDicke1961,dicke1962physical,brans1962mach} with a constant vacuum energy density, the so-called BD-$\Lambda$CDM \cite{Avilez:2013dxa,deCruzPerez:2018cjx,SolaPeracaula:2019zsl,SolaPeracaula:2020vpg,Joudaki:2020shz,SolaPeracaula:2021gxi}. The {\it Planck} collaboration already studied the impact of marginalization in the concordance model, using the 2013 likelihood \cite{Planck:2013nga}. They did not find any significant difference between the results obtained with the profile likelihoods and those found with the marginalized posteriors, basically because the non-marginalized PDF for the $\Lambda$CDM parameters is very close to a multivariate Gaussian. We perform our study using the 2018 {\it Planck} likelihood \cite{Planck:2018vyg}, as a baseline analysis to test our method. We confirm {\it Planck}'s results, as expected, also when we add data on supernovae of Type Ia (SNIa) and baryon acoustic oscillations (BAO). Then we apply our methodology to the other models, using again the CMB+SNIa+BAO combined dataset. For EDE we find a result close to the one reported recently in \cite{Herold:2021ksg}, in which the authors used the PD method to constrain the maximum EDE fraction, $f_{\rm ede}$, and found $f_{\rm ede}= 0.072\pm 0.036$ at $68\%$ c.l. using the {\it Planck} 2018 data and the BOSS DR12 full-shape likelihood. Their study showed that volume effects play actually a role in the determination of the confidence intervals of $f_{\rm ede}$, something that was already hinted in \cite{Smith:2020rxx}. This allows us to understand why the authors of \cite{Ivanov:2020ril} found no evidence for EDE using the same dataset. They obtained $f_{\rm ede}<0.072$ at $95\%$ c.l. using the marginalization technique, with no peak in the posterior, cf. Fig. 9 in \cite{Ivanov:2020ril}. In this work we compare explicitly the results obtained with the MC using both, the marginalization and the PD approaches. Using the profile distribution we find a similar (although somewhat tighter) result for $f_{\rm ede}$ to the one in \cite{Herold:2021ksg}, $f_{\rm ede}\simeq 0.052^{+0.022}_{-0.021}$ at $68\%$ c.l., whereas with the marginalization we obtain $f_{\rm ede}<0.048$ at $68\%$ c.l. Getting rid of the volume effects, we observe a peak for the EDE fraction, $\sim 2.5\sigma$ away from 0, similar to \cite{Herold:2021ksg}. We also discuss for the first time in the literature the PD constraints on the other parameters of the theory. We observe $\sim 1\sigma$ shifts in the PDs of some parameters, e.g. $H_0$, when compared to the marginalized posteriors. We discuss the status of the cosmological tensions in the light of these results. Finally, we perform the marginalization and PD analyses also for the CDE and BD-$\Lambda$CDM models. We do not find in these cases very significant differences between these two methods. Our results are obtained directly from the Markov chains, so they are not as accurate as the ones we would get doing a dedicated PD analysis. However, they already inform us about the existing marginalization biases and the shifts we expect to obtain in the constraints when performing a more precise study with the PDs. And most importantly, they are basically for free, since the PDs can be estimated very fast from the Markov chains. 

We provide our conclusions in Sec. \ref{sec:conclusions}.


\section{A trivial example in 2D}\label{sec:SecI}

We dedicate this section to illustrate the problem discussed in the Introduction through a very simple example, in a two-dimensional parameter space composed by the pair of dimensionless parameters $\alpha$ and $\beta$. We consider the following normalized probability distribution, built from the sum of two bivariate Gaussians,

\begin{equation}\label{eq:exampleDist}
\mathcal{P}(\alpha,\beta)=\frac{5}{21\pi}[e^{-\frac{1}{4}(\alpha^2 + \beta^2)} + e^{-\frac{1}{4}(\alpha- 3.5)^2-100\beta^2}]\,.
\end{equation} 
The origin of this distribution is not important for the discussion, but we can imagine that it results from an experiment that we can call {\it experiment 1}. The latter is used to constrain $\alpha$ and $\beta$, which are the parameters of some theory we want to test. In Fig. \ref{fig:fig1} we show its shape and the corresponding confidence regions at $68\%$ and $95\%$ c.l. in the $(\alpha,\beta)$-plane. A huge fraction of the total volume under this distribution is below the first term of \eqref{eq:exampleDist}, in the region $\alpha^2+\beta^2\lesssim 4$. This can be appreciated in the left plot of Fig. \ref{fig:fig1}. Notice, though, that there is a small volume centered at $\beta=0$ that comes out from the latter, which extends up to larger values of $\alpha$, namely up to $\alpha\sim 6$. This is due to the second term in \eqref{eq:exampleDist}, which has a very small covariance for $\beta$. As a result, the contours in the $(\alpha,\beta)$-plane acquire a peculiar ``Pinocchio'' shape (see the right plot in Fig. \ref{fig:fig1}). Thus, it is clear from the two plots in that figure that it is possible to find points in parameter space with $\alpha\sim 6$ leading to high values of the probability density \eqref{eq:exampleDist}, provided that $\beta$ is close enough to $0$.

\begin{figure*}[t!]
\begin{center}
\includegraphics[width=6.5in, height=2.5in]{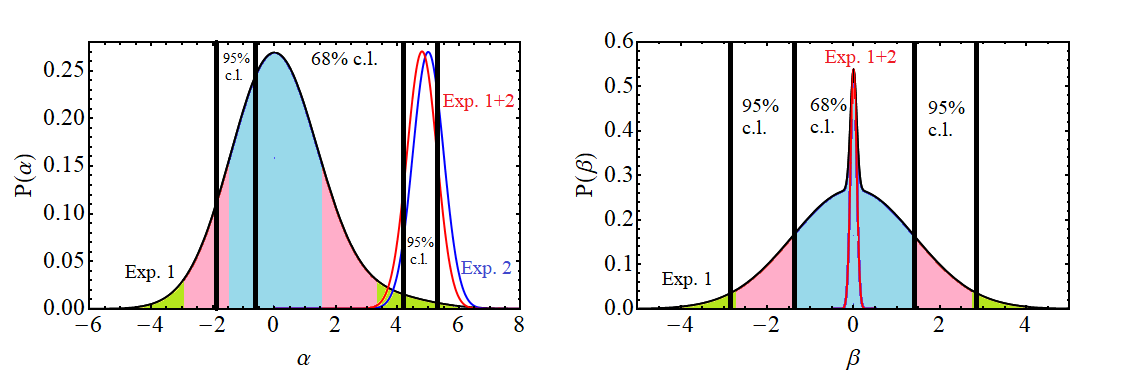}
\caption{Marginalized one-dimensional posterior distributions for $\alpha$ and $\beta$ obtained from {\it experiment 1}, in black, cf. formulae \eqref{eq:posteriorAlpha} and \eqref{eq:posteriorBeta}. The cyan, pink and green areas below them indicate the $68\%$, $95\%$ and $99.7\%$ c.l. regions, respectively, obtained from the marginalized posteriors. The black thick vertical lines are the borders of the $68\%$- and $95\%$-regions obtained with the PD method. The curve in blue of the left plot is the Gaussian constraint on $\alpha$, obtained from {\it experiment 2}. It is in $\sim 3\sigma$ tension with the marginalized result from {\it experiment 1}, whereas it is compatible at $\sim 1\sigma$ c.l. with the PD result. The volume effects introduced in the marginalization process explain this difference. The marginalized posteriors of $\alpha$ and $\beta$ derived from the joint constraint from {\it experiments 1} and {\it 2} are shown in red in the two plots as well. Both  experiments can be safely combined, which allows us to obtain much tighter constraints on the two parameters.}\label{fig:fig2}
\end{center}
\end{figure*}

Let us consider now a second experiment, called {\it experiment 2}, which is able to constrain only $\alpha$: $\alpha=5.0\pm 0.5$ at $1\sigma$ c.l., with a probability distribution that in this case we consider to be Gaussian, for simplicity. $\beta$ is unconstrained by {\it experiment 2}. It is evident that {\it experiments} {\it 1} and {\it 2} are compatible at $\sim 1\sigma$ c.l. The $1\sigma$-regions of the contours of the two experiments in the $(\alpha,\beta)$-plane overlap, and $\alpha=5$ is close to one of the peaks of the distribution \eqref{eq:exampleDist}, see again the plots in Fig. \ref{fig:fig1}. The protuberance at $\beta\sim 0$ in the confidence contours from {\it experiment 1} is what makes possible such a good agreement.

Now we can ask ourselves what happens if we study the compatibility of the two experiments making use of the marginalized posteriors, instead of \eqref{eq:exampleDist}. Will we come to the same conclusion? The marginalized one-dimensional posteriors can be obtained analytically using \eqref{eq:marg} and \eqref{eq:exampleDist}. They read, 

\begin{equation}\label{eq:posteriorAlpha}                           
\mathcal{P}(\alpha) = \frac{5}{21\sqrt{\pi}}\left[2e^{-\frac{\alpha^2}{4}}+0.1e^{-\frac{1}{4}(\alpha-3.5)^2}\right]\,,
\end{equation}

\begin{equation}\label{eq:posteriorBeta}  
\mathcal{P}(\beta) = \frac{10}{21\sqrt{\pi}}\left[e^{-\frac{\beta^2}{4}}+e^{-100\beta^2}\right]\,.
\end{equation}
We show their shape in Fig. \ref{fig:fig2}, indicating with different colors the $68\%$, $95\%$ and $99.7\%$\footnote{In several parts of this manuscript we express the confidence intervals in terms of the number of sigmas, which is not completely equivalent: $1\sigma=68.23\%$ c.l., $2\sigma=95.44$ c.l., and $3\sigma=99.74\%$ c.l.} confidence intervals derived from them. One can see in the left plot that the point $\alpha=5$, i.e. the preferred value by {\it experiment 2}, falls now almost entirely inside the $3\sigma$ region of {\it experiment 1}. Therefore, by looking at the marginalized one-dimensional posterior distribution for $\alpha$ we would conclude that there is a $\sim 3\sigma$ tension between {\it experiments 1} and {\it 2}. We have already seen, though, that this tension is completely non-existent. It is only an artifact introduced by the marginalization process. This example illustrates in a crystal-clear way how, if not interpreted correctly, the marginalized distributions can make us extract wrong conclusions. For instance, by looking at the left plot of Fig. \ref{fig:fig2} we could conclude that there is an unaccounted systematic error affecting one of the two (or even both) experiments, or that the theoretical framework used to describe them is just insufficient. In reality, none of this is needed at all, since the results obtained from the two experiments are fully compatible. It is clear from the right plot in Fig. \ref{fig:fig1} that they are consistent with each other and therefore can be safely employed together to strongly improve the constraints on the parameters of the underlying theory. By doing so we obtain the red contours in the right plot of Fig. \ref{fig:fig1}, and the posteriors in red of Fig. \ref{fig:fig2}.

The one-dimensional marginalized distribution of a particular parameter do not tell us directly what values of that parameter fit better the data, but the probability that a particular value of that parameter explains the data. These two things are not equivalent. There is a subtle but crucial difference. The latter is computed basically by ``summing'' the contribution of all points in parameter space given a fixed value of the parameter of interest, obtaining in this way an integrated probability. This introduces the so-called ``volume effects''. They are the reason why some values of parameters that are able to fit better or equally well the data can be underrated in the marginalized distribution, just because the volume in parameter space occupied by  the points that lead to such a good description of the data for that particular value of the parameter of interest is smaller. If there are values of that parameter that can fit very well the data, but in too small regions of parameter space, the marginalized distribution will in general hide them. Therefore we can lose precious information in the marginalization process. This is what happens in our example. Given two values of $\alpha$ leading to similar values of the original distribution \eqref{eq:exampleDist} (for some concrete values of $\beta$), let us say e.g. $\alpha=0$ and $\alpha=5$, the one with an associated larger volume containing more points with high $\mathcal{P}(\alpha,\beta)$ will lead to a larger value of the posterior distribution $\mathcal{P}(\alpha)$. In this case, $\alpha=0$ will be much more favored than $\alpha=5$ due to the existence of more points in parameter space with large $\mathcal{P}(\alpha,\beta)$ in the line $\alpha=0$ than in line $\alpha=5$.

\begin{figure*}[t!]
\begin{center}
\includegraphics[width=6.5in, height=2.3in]{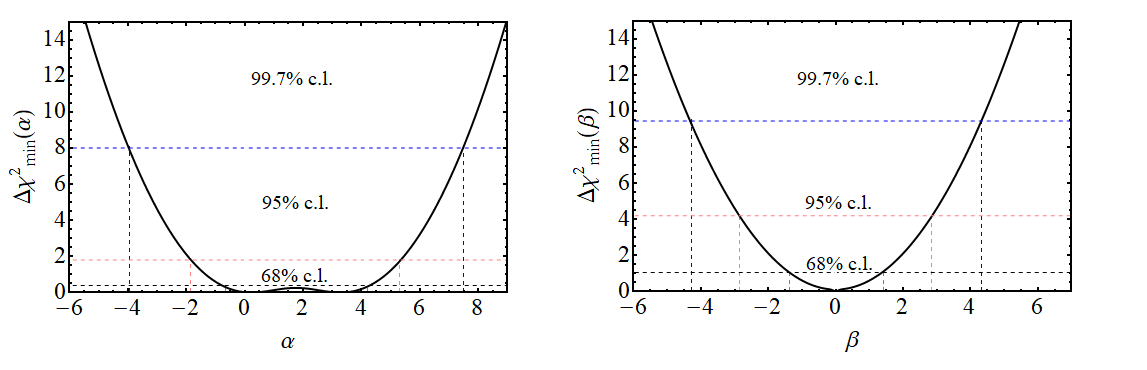}
\caption{$\Delta\chi^2_{\rm min}$ obtained from {\it experiment} 1 as a function of $\alpha$ and $\beta$, respectively. The horizontal dashed lines indicate the values of $\Delta\chi^2_{\rm min}$ that set the borders between the $68\%$ (in black), $95\%$ (in pink) and $99.7\%$ (in blue) c.l. regions. The vertical dashed lines indicate the values of the parameters associated to these borders.}\label{fig:fig3}
\end{center}
\end{figure*}   

This example is trivial, of course, since we are dealing only with two parameters. It has served us just to illustrate the problem. The detection of biases in the conclusions that we extract from the marginalized posteriors is easy in this case because we can visualize the results from the original (non-marginalized) distribution \eqref{eq:exampleDist} by plotting it together with the confidence contours in the plane $(\alpha,\beta)$, which is already the full parameter space of the theory. Detecting biases introduced by the marginalization step can be much more difficult in parameter spaces with a larger number of dimensions because we cannot make plots representing the non-marginalized distribution when we have three or more parameters. The information loss in the marginalization process can be huge in these cases, and we cannot check its impact by only having a look to the marginalized one- and two-dimensional posterior distributions. We clearly need to proceed in an alternative way, which avoids the marginalization step and works directly with the original distribution.

We suggest to plot the maximum value of the original distribution \eqref{eq:exampleDist} as a function of the value of the parameter we are interested in. Alternatively, we can build the plot of the minimum of $\Delta\chi^2=-2\ln(\mathcal{P}/\mathcal{P}_{\rm max})$, $\Delta\chi^2_{\rm min}$, as a function of the parameter of interest, with $\mathcal{P}_{\rm max}$ the maximum value of \eqref{eq:exampleDist}. In our example, we can draw the plots for both, $\alpha$ and $\beta$, i.e. $\Delta\chi^2_{\rm min}(\alpha)=\chi^2_{\rm min}(\alpha)-\chi^2_{\min}$ and $\Delta\chi^2_{\rm min}(\beta)=\chi^2_{\rm min}(\beta)-\chi^2_{\min}$, where $\chi^2_{\min}$ is the absolute minimum value of the $\chi^2$. For every fixed $\alpha$ we have a model that depends only on $\beta$. The plot of $\Delta\chi_{\rm min}^2(\alpha)$ compares the fitting performance of these models to the model with lowest $\chi^2$. The quantity $\Delta\chi_{\rm min}^2(\alpha)$ can be used as an information criterion. Since we are comparing models with the same number of parameters constrained with the same number of data points, $\Delta\chi_{\rm min}^2(\alpha)$ is actually equivalent to the differences $\Delta {\rm AIC}(\alpha)={\rm AIC}(\alpha)-{\rm AIC}_{\rm min}$ and $\Delta {\rm BIC}(\alpha)={\rm BIC}(\alpha)-{\rm BIC}_{\rm min}$, with AIC and BIC referring to the Akaike and Bayesian (or Schwarz) information criteria \cite{Akaike,Schwarz1978,KassRaftery1995}, respectively\footnote{Recall that for a given model $\mathcal{M}$, ${\rm AIC}(\mathcal{M})=\chi^2_{\rm min}(\mathcal{M})+\frac{2kn}{n-k-1}$ and ${\rm BIC}(\mathcal{M})=\chi^2_{\rm min}(\mathcal{M})+k\ln(n)$, with $k$ the number of fitting parameters and $n$ the number of data points.}. The same correspondence exists between $\Delta\chi_{\rm min}^2(\beta)$, $\Delta {\rm AIC}(\beta)$ and $\Delta {\rm BIC}(\beta)$, of course. This allows us to assign a probability weight to each model, i.e. to each value of the parameter we are interested in. For instance, for $\alpha$ it can be done as follows, 

\begin{equation}\label{eq:weight}
w(\alpha)=\frac{e^{-\chi^2_{\rm min}(\alpha)/2}}{\int e^{-\chi^2_{\rm min}(\alpha)/2}\,d\alpha}=\frac{\tilde{\mathcal{P}}(\alpha)}{\int \tilde{\mathcal{P}}(\alpha)\,d\alpha}\,,
\end{equation}
with $\tilde{\mathcal{P}}$ the profile distribution as defined in \eqref{eq:PD}. The integrals are performed over the range $\alpha\in(-\infty,+\infty)$. Dividing the numerator and denominator of the last expression by $\tilde{\mathcal{P}}_{\rm max}$ we get

\begin{equation}\label{eq:weight}
w(\alpha)=\frac{R(\alpha)}{\int R(\alpha)\,d\alpha}\,,
\end{equation}
where $R$ is the ratio \eqref{eq:ratio}. We have already normalized the weights such that 

\begin{equation}
\int w(\alpha)\,d\alpha = 1\,.
\end{equation}
The analogous quantities for $\beta$ can be computed straightforwardly. Using \eqref{eq:weight} we can obtain the confidence intervals, without introducing any volume effect. For instance, to determine the $68\%$ c.l. interval for $\alpha$ we proceed as follows. We first find the values of $\alpha_1$ and $\alpha_2$ such that

\begin{equation}\label{eq:cond}
\int_{\alpha_1}^{\alpha_2}w(\alpha)\,d\alpha=0.68\,,\quad {\rm with}\quad w(\alpha_1)=w(\alpha_2)\,.
\end{equation}
In practice, for every value of $w$ we can obtain numerically the associated values of $\alpha$. We solve the resulting equation starting from the maximum value $w_{\rm max}$, and repeat this calculation with decreasing $w$ until fulfilling the condition \eqref{eq:cond}. In the example under study there is no physical boundary hitting any of the integration limits. If there was e.g. a lower physical boundary at $\alpha_b$ such that

\begin{equation}
\int_{\alpha_b}^{\tilde{\alpha}_b}w(\alpha)\,d\alpha<0.68\,,\quad {\rm with}\quad w(\tilde{\alpha}_b)=w(\alpha_b)\,,
\end{equation}
we would fix $\alpha_1=\alpha_b$ and would vary only the upper limit of the integral, until obtaining the desired area (confidence level) below that fraction of the curve $w$ \cite{Feldman:1997qc}. The analogous procedure would be applied if there was a physical boundary in the upper range, of course. For multimodal distributions one can also apply this method, and can obtain in general disjoint confidence regions. 

Let us apply now this approach to the example under study. In Fig. \ref{fig:fig3} we show $\Delta\chi^2_{\rm min}$ obtained from {\it experiment} 1 as a function of $\alpha$ and $\beta$ in the left and right plots, respectively, together with the corresponding borders of the $68\%$, $95\%$ and $99.7\%$ c.l. regions, computed as explained in the previous paragraphs. The left plot of Fig. \ref{fig:fig3} shows that the $68\%$ c.l. region covers the range $-0.65\leq\alpha\leq 4.20$, whereas the $95\%$ c.l. region extends up to $-1.87\leq\alpha\leq 5.30$. We show these borders explicitly in the left plot of Fig. \ref{fig:fig2} as well. By looking at it one can see that these borders are at odds with the $68\%$-region obtained from the marginalized posterior of $\alpha$ \eqref{eq:posteriorAlpha}. The latter encompasses a much smaller range of values ($-1.46\leq\alpha\leq 1.55$) and points to a $\sim 3\sigma$ tension with {\it experiment 2}, as already mentioned before. Applying the PD method we are able to extract the correct conclusion, i.e. that {\it experiments 1} and {\it 2} are compatible at $\sim 1\sigma$ c.l. The constraint from {\it experiment 2} is well within the ``consistency'' region derived with the profile distribution, and so is also the joint constraint from {\it experiments 1} and {\it 2} (cf. again Fig. \ref{fig:fig2}). For $\beta$, instead, the constraints obtained with the PD and marginalization methods are quite similar.

The marginalized posteriors give us valuable information. They inform us about the regions in parameter space that contain the bulk of the probability. There can be, though, some (smaller) regions containing points able to explain equally well (or even better) the data, which are hidden in the marginalized posteriors. This can be a problem when we want to study tensions between observations, between models, or between models and observations. The PD method allows us to detect fictitious tensions that are induced by volume effects coming from the marginalization procedure. It is complementary, in the sense that it can be used together with the marginalization method to better interpret the results we get from the Monte Carlo analyses. They provide different and supplementary statistical information. In the example under consideration, by looking to the results presented in Fig. \ref{fig:fig2}, we can make the following statements combining the information gathered with the PD and marginalization methods: (i) the $68\%$ of the probability is concentrated in the regions $-1.46\leq\alpha\leq 1.55$ (for $\alpha$) and $-1.34\leq\beta\leq 1.37$ (for $\beta$). This is what the one-dimensional marginalized posteriors tell us; (ii) nevertheless, it is still possible to find regions in parameter space satisfying $-2.0\leq\alpha\leq 5.5$ (for $\alpha$) and $-1.9\leq\beta\leq 1.9$ (for $\beta$) that lead to a good description of the data, close to the one offered by the best-fit point. They are not detected by the marginalized posterior, just because they occupy a volume in parameter space that is too small and are penalized when compared to other regions with a larger integrated probability. Thus, the apparent tension observed in the marginalized posterior for $\alpha$ does not exist, and the two experiments can be combined without problems. 

It is clear that volume effects can leave an imprint on the conclusions we extract from MC analyses. Assessing its precise impact with the marginalized posteriors is not possible, and this is even more complicated in high-dimensional parameter spaces, as those we usually deal with in cosmological studies. In the next section we analyze the impact of marginalization on several cosmological models comparing the results obtained with the marginalization and PD methods, in the light of modern data.


\section{Volume effects in cosmology}\label{sec:SecII}

\subsection{Data}\label{sec:data}

In this work we use the full {\it Planck} 2018 TTTEEE+lowE+lensing likelihood (Planck18, in short) \cite{Planck:2018vyg}, which incorporates the data on the temperature (TT) and polarization (EE) anisotropies of the cosmic microwave background (CMB), and also their cross-correlations (TE) at low and high multipoles, together with the CMB lensing likelihood. We vary in our MCs the 21 {\it Planck} nuisance parameters employed to model the experiment systematics. 

We also consider the data on the apparent magnitudes and redshifts from the standardized 1048 SNIa of the Pantheon compilation \cite{Scolnic:2017caz}, taking into account the statistical and systematic uncertainties through the corresponding covariance matrix. The absolute magnitude of the SNIa, $M$, is left free in the MC runs.

We also use the following BAO data in combination with the CMB and SNIa likelihoods: 

\begin{itemize}

\item $D_V/r_d$ at $z=0.122$ provided in \cite{Carter:2018vce}, which combines the dilation scales previously reported by the 6dF Galaxy Survey (6dFGS) \cite{Beutler:2011hx} at $z=0.106$ and the Sloan Digital Sky Survey (SDSS) Main Galaxy Sample at $z=0.15$ \cite{Ross:2014qpa}. The dilation scale $D_V$ reads,

\begin{equation}
D_V(z)=\left[D_M^2(z)\frac{cz}{H(z)}\right]^{1/3}\,,
\end{equation}
with $D_M=(1+z)D_{A}(z)$ the comoving angular diameter distance. The distance $r_d$ is the sound horizon at the baryon drag epoch. 

\item The anisotropic BAO data ($D_A(z)/r_d$, $H(z)r_d$) measured by BOSS using the LOWZ ($z=0.32$) and CMASS ($z=0.57$) galaxy samples \cite{Gil-Marin:2016wya}.

\item The dilation scale measurements by WiggleZ at $z=0.44,0.60,0.73$ \cite{Kazin:2014qga}.

\item $D_A(z)/r_d$ at $z=0.81$ measured by the Dark Energy Survey (DES) \cite{DES:2017rfo}.

\item  The anisotropic BAO data from the extended BOSS Data Release 16 (DR16) quasar sample at $z=1.48$ \cite{Neveux:2020voa}.

\item The anisotropic BAO information obtained from the Ly$\alpha$ absorption and quasars of the final data release (SDSS DR16) of eBOSS, at $z=2.334$ \cite{duMasdesBourboux:2020pck}.

\end{itemize}
The combination of CMB and BAO allows us to construct the inverse cosmic distance ladder \cite{Aubourg:2014yra,Cuesta:2014asa,Feeney:2018mkj,Camarena:2019rmj}, which is relevant for the discussion on the $H_0$ tension. On the other hand, the SNIa data let us tighten the constraints on the cosmological parameters and constrain also the absolute magnitude of the SNIa. The large value of $M$ measured by the SH0ES collaboration \cite{Riess:2021jrx} using the calibration of the SNIa in the first steps of the cosmic distance ladder can be considered to be the reason why the value of $H_0$ that is obtained with the SNIa in the Hubble flow (using the calibrated value of $M$) is much larger than the one preferred by the {\it Planck} data, under the assumption of the concordance model, see e.g. \cite{Camarena:2019moy,Efstathiou:2021ocp}. Therefore, it is interesting to study the constraints on $M$ that are obtained with the various cosmological models. We display its value in all our tables.


\begin{figure*}[t!]
\begin{center}
\includegraphics[width=6.5in, height=5.8in]{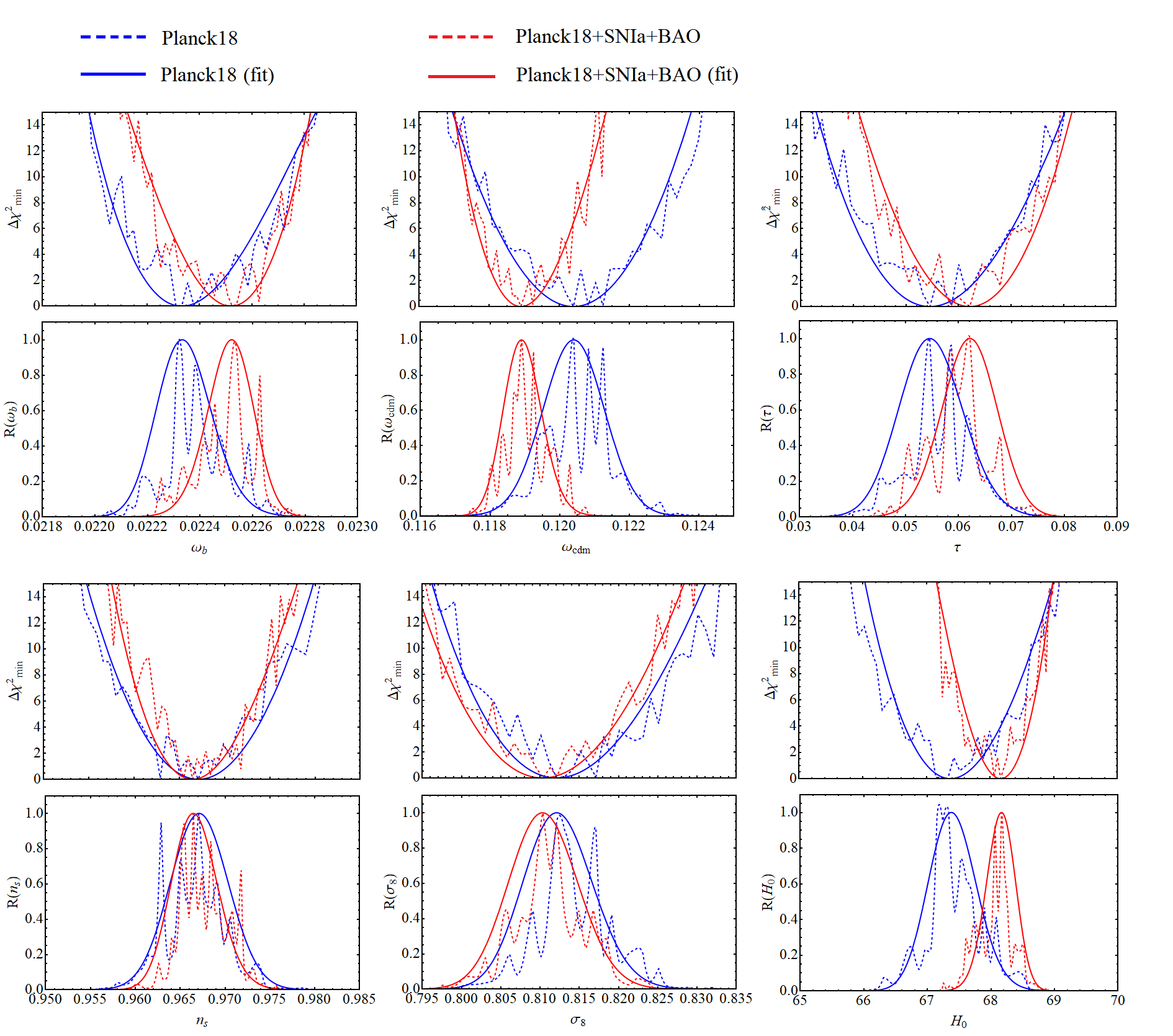}
\caption{Plots of $\Delta\chi^2_{\rm min}(p_i)$ and $R(p_i)=\exp{(-\Delta\chi^2_{\rm min}(p_i)/2)}$ for the $\Lambda$CDM parameters contained in the vector $\vec{p}=(\omega_{b},\omega_{\rm cdm},\tau,n_s,\sigma_8,H_0)$, with $H_0$ expressed in km/s/Mpc. We include both, the noisy result obtained from the MC Markov chains (solid lines) and the smoothed one obtained by making use of the fitting formula \eqref{eq:fit1} (dashed lines). We show the results from the datasets Planck18 (in blue) and Planck18+SNIa+BAO (in red), see Secs. \ref{sec:data} and \ref{sec:LCDM}, and Table I.}\label{fig:fig4}
\end{center}
\end{figure*}

\subsection{Methodology}\label{sec:methodology}

The main aim of this paper is to show how to assess the impact of the volume effects using the Markov chains obtained in MC studies, the same ones from which one usually gets the marginalized results. This can tell us whether a dedicated PD analysis is needed and, if so, for what parameters in particular. This can be done straightforwardly and in a very fast way, and allow us to build a clearer picture of the biases that are introduced by the marginalization step.

\begin{table*}[t!]
\centering
\begin{tabular}{|c ||c | c || c| c|    }
 \multicolumn{1}{c}{} & \multicolumn{1}{c}{} & \multicolumn{1}{c}{} & \multicolumn{1}{c}{} & \multicolumn{1}{c}{} \\
\multicolumn{1}{c}{$\Lambda$CDM} &  \multicolumn{2}{c}{Planck 2018} &  \multicolumn{2}{c}{Planck18+SNIa+BAO}
\\\hline
{\small Parameter} & {\small Marginalization}  & {\small PD}& {\small Marginalization} & {\small PD}
\\\hline
$\omega_b$ & $0.02238\pm 0.00015$ & $0.02233^{+0.00014}_{-0.00013}$ & $0.02247^{+0.00013}_{-0.00014}$ & $0.02252^{+0.00010}_{-0.00012}$ \\\hline
$\omega_{\rm cdm}$ & $0.1204^{+0.0012}_{-0.0013}$ & $0.1204\pm 0.012$ & $0.1190\pm 0.0008$ & $0.1189\pm 0.0007$ \\\hline
$n_s$ & $0.966\pm 0.004$ & $0.967\pm 0.004$ & $0.968\pm 0.004$ & $0.966\pm 0.004$ \\\hline
$\tau$ & $0.055^{+0.007}_{-0.008}$ & $0.055 \pm 0.008$ & $0.058^{+0.007}_{-0.008}$ & $0.062\pm 0.007$ \\\hline
$\sigma_8$ & $0.814\pm 0.006$ & $0.812^{+0.006}_{-0.007}$ & $0.811\pm 0.006$ &  $0.810^{+0.007}_{-0.006}$ \\\hline
$H_0$ [km/s/Mpc] & $67.4\pm 0.6$ & $67.4\pm 0.6$ & $68.0\pm 0.4$ & $68.1\pm 0.4$ \\\hline\hline
$r_d$ [Mpc] & $146.92\pm 0.27$ & $147.01\pm 0.26$ & $147.18^{+0.20}_{-0.22}$ & $147.17\pm 0.20$ \\\hline
$M$ & $-$ & $-$ & $-19.408\pm 0.010$ & $-19.403\pm 0.010$ \\\hline
$S_{8}$ & $0.832^{+0.013}_{-0.014}$ & $0.833^{+0.013}_{-0.012}$ & $0.818^{+0.010}_{-0.009}$ & $0.816\pm 0.009$ \\\hline
$\sigma_{12}$ & $0.807^{+0.008}_{-0.009}$ & $0.808^{+0.008}_{-0.009}$ & $0.799\pm 0.007$ & $0.798\pm 0.007$ \\\hline
$S_{12}$ & $0.814^{+0.010}_{-0.011}$  & $0.813^{+0.010}_{-0.009}$  & $0.803\pm 0.008$ & $0.801^{+0.008}_{-0.007}$ \\\hline
\end{tabular}
\label{tab:table}                          
\caption{Comparison between the constraints obtained with the marginalization and PD methods for the $\Lambda$CDM, using the Planck2018 and Planck2018+SNIa+BAO datasets. For the marginalization method we list the mean and $1\sigma$ uncertainties of the main cosmological parameters, together with some derived and nuisance parameters of interest. For the PD method, instead, we show the values at the peak of the one-dimensional PDs, and the corresponding $1\sigma$ uncertainties. Both methods lead to very similar results, cf. Sec. \ref{sec:LCDM} for details.}
\end{table*}

The procedure is very simple and can be applied to the analysis of any cosmological model. The steps are as follows. First, we perform the MC analysis. We use our own modified version of \texttt{CLASS} \cite{Blas:2011rf} to solve the Einstein-Boltzmann equations for each model, and run the MCs with \texttt{MontePython} \cite{Audren:2012wb}, considering the datasets described in Sec. \ref{sec:data}. We use the Metropolis-Hastings algorithm \cite{Metropolis:1953aaa,Hastings:1970bbb}. For the $\Lambda$CDM and BD$-\Lambda$CDM models we stop the MC when the Gelman-Rubin convergence parameter \cite{GelmanRubin1992} $R-1<0.01$, whereas for ULA and CDE, which have a slower convergence, we stop the MC when $R-1<0.05$. The acceptance rate is between $32\%$ and $40\%$ in all our MC runs. From the resulting Markov chains we can get the marginalized posteriors, and also can draw the plots of $\Delta\chi^2_{\rm min}$ (and the ratio $R$ \eqref{eq:ratio}) for all the parameters. This can be done very efficiently, just binning the range between the minimum and maximum values of the parameters and selecting the minimum $\chi^2$ in each bin. The quantity $\chi^2_{\rm min}$ is obviously the minimum value in the list. This information can be directly read from the Markov chains. Due to the limited number of points in the chains, we expect, of course, the resulting plots of $\Delta\chi^2_{\rm min}$ to have some noise. We do not pretend (nor expect) to be extremely accurate at this point. As already mentioned, our goal is to detect biases and estimate their impact on the various parameters. The efficiency of this method allows us to apply it not only to the main cosmological parameters, but also to the derived and nuisance parameters, which is also positive. Once we build these plots we can obtain smoothed curves using some fitting or envelope formula and compute the uncertainties of the parameters with the PD method, using \eqref{eq:weight}, exactly as we have done in the example of Sec. \ref{sec:SecI}.  

In the next section we apply this methodology to the study of four different cosmological models and discuss the impact of marginalization on each of them.


\subsection{Results}\label{sec:results}

In this section we present the results for the $\Lambda$CDM (Sec. \ref{sec:LCDM}), EDE (Sec. \ref{sec:EDE}), CDE (Sec. \ref{sec:CDE}) and the BD-$\Lambda$CDM (Sec. \ref{sec:BDLCDM}). We consider in all cases a perturbed flat Friedmann-Lema\^itre-Robertson-Walker (FLRW) spacetime, and a massive neutrino of $0.06$ eV. We do not perform an exhaustive study of the physics behind each model, since this has been already done in dedicated papers in the literature, cf. the corresponding references in each section. Here we just want to compare the results obtained with the marginalization and PD methods, obtained directly from the MC Markov chains, see whether volume effects play a role or not and, if so, quantify their impact. 

\subsubsection{$\Lambda$CDM}\label{sec:LCDM}

We present our main results for the concordance model in Fig. \ref{fig:fig4} and Table I. In Fig. \ref{fig:fig4} we show the plots of $\Delta\chi^2_{\rm min}$ obtained from the MC Markov chains for the six main parameters of the model, namely, the reduced density parameters for baryons and cold dark matter (CDM), $\omega_{b}$ and $\omega_{\rm cdm}$, the reionization depth, $\tau$, the spectral index of the primordial power spectrum, $n_s$, the {\it rms} of mass fluctuations at scales of $R_8=8h^{-1}$ Mpc, $\sigma_8$, and the Hubble parameter, $H_0$. We use the Planck2018 and Planck2018+SNIa+BAO datasets. The curves in Fig. \ref{fig:fig4} are a bit noisy for the reason explained in Sec. \ref{sec:methodology}. We smooth them using the fitting formula 

\begin{equation}\label{eq:fit1}
\Delta\chi^2_{\rm min}(p)= a(p-p_{\rm bf})^2+b(p-p_{\rm bf})^3\,,
\end{equation}  
where $p$ is the parameter of interest, $p_{\rm bf}$ is the value leading to $\Delta\chi^2_{\rm min}(p_{\rm bf})=0$, and $a$ and $b$ are the fitting coefficients, whose values obviously depend on the parameter under study. The second term in the {\it rhs} of \eqref{eq:fit1} allows for some departure from Gaussianity. We also include the shape of the fitting formula in Fig. \ref{fig:fig4}, for comparison, and the plots of the ratios \eqref{eq:ratio}, $R(p)=\exp(-\Delta\chi^2_{\rm min}(p)/2)$. The fitting formula \eqref{eq:fit1} naturally produces an underestimation of the parameter uncertainties due to the fact that it is not an exact envelope of the PD. This can be compensated in an empirical (approximate) way by considering an additional contribution to the uncertainties equal to twice the size of the bins, which is the typical period of the oscillations caused by the noise. In Table I we compare the results obtained with the marginalization and PD methods. There are no significant differences between them, since the PDF for the $\Lambda$CDM parameters is very close to a multivariate Gaussian. Our results resonate perfectly well with those found by the {\it Planck} collaboration using the 2013 likelihood \cite{Planck:2013nga}. Notice that we also show the results for the {\it rms} mass fluctuation in spheres of radius $12$ Mpc, $\sigma_{12}$, and the related quantity $S_{12}=\sigma_{12}(\omega_{m}/0.14)^{0.4}$, which according to \cite{Sanchez:2020vvb} characterize better the amplitude of the linear matter power spectrum than the usually employed $\sigma_8$ and $S_8=\sigma_8(\Omega_m/0.3)^{0.5}$, since the scale at which they are computed does not depend on $H_0$. Changes in $\sigma_{12}/S_{12}$ are due to changes in the shape of the power spectrum, not to shifts of the scale at which the {\it rms} of mass fluctuations is computed. As reported in \cite{Gomez-Valent:2021cbe}, in some models there can be non-negligible differences between the statistical tension with the large-scale structure (LSS) data when it is quantified with $\sigma_{12}/S_{12}$ instead of $\sigma_8/S_8$. For instance, in the $\Lambda$CDM the tension seems to be a bit lower in terms of $\sigma_{12}/S_{12}$, although this is difficult to judge in a precise way, since weak lensing and galaxy surveys do not provide constraints on these quantities. We show these values for completeness in some of our tables, together with the absolute magnitude of SNIa, $M$, and the sound horizon, $r_d$. The latter help us to understand the status of the $H_0$ tension in the context of the various models under study. Volume effects have no impact on the cosmological tensions in the $\Lambda$CDM, since marginalization does not introduce any important shift in the central values and uncertainties of its cosmological parameters.


\subsubsection{Early dark energy}\label{sec:EDE}

We consider a self-conserved scalar field minimally coupled to gravity with the following potential,

\begin{equation}\label{eq:potentialEDE}
V(\phi)=V_0+m^2f^2\left[1-\cos\left(\frac{\phi}{f}\right)\right]^3\,,
\end{equation}
with $m$ and $f$ being the two parameters of the theory (both with dimensions of mass in natural units), typically satisfying $m^2f^2\gg V_0$, and $V_0$ close to the value of the energy density associated to the cosmological constant $\Lambda$ in the standard model. They have to be considered together with the six $\Lambda$CDM parameters and the initial condition for the scalar field, usually expressed as $\theta_{\rm ini}\equiv \phi_{\rm ini}/f$. The modified Klein-Gordon equation,

\begin{equation}
\ddot{\phi}+3H\dot{\phi}+\frac{\partial V}{\partial\phi}=0\,,
\end{equation}
governs the dynamics of the scalar field, with the dot denoting a derivative with respect to the cosmic time. Deep in the radiation-dominated epoch, when the expansion rate of the universe is much larger than the mass of the scalar, i.e. when $H\gg m$, the scalar field is frozen and the potential remains basically constant. Hence, we can safely take the initial condition $\dot{\phi}_{\rm ini}=0$. The scalar field starts to evolve when $H\sim m$. At that time it rolls down the potential and eventually oscillates around the minimum, making its energy density to decay faster than radiation. After the decay the model simply reduces to the concordance model, with $V(\phi)\approx V_0$. 

This EDE model has been intensively studied in the last years as a possible solution to the Hubble tension \cite{Poulin:2018dzj,Poulin:2018cxd}. If the constant value of the potential during the radiation-dominated era is sufficiently large, and if the parameter $m$ is tuned such that $m\sim H(z_{eq})$, with $z_{eq}$ the redshift at the matter-radiation equality time, one can produce a peak in the EDE fraction close to $z_{eq}$, at $z_{\rm max}$, that can give rise to a decrease of the sound horizon $r_d$ and lead, in turn, to larger values of the Hubble function in the late-time universe, which is required to keep the position of the first peak of the CMB temperature anisotropies and the good description of the BAO data. The presence of EDE before the decoupling of the CMB photons produces an enhancement of the early integrated Sachs-Wolfe (iSW) effect, which has to be compensated by an increase of $\omega_{\rm cdm}$ and $n_s$, see e.g. \cite{Vagnozzi:2021gjh}. For a fixed $V_0$, this increases the LSS in the universe due to the more efficient growth of the matter fluctuations in the matter- and $\Lambda$-dominated eras. The extent up to what this fact might hinder the ability of the model to loosen the $H_0$ tension has been already studied in the literature, see the works \cite{Hill:2020osr,DAmico:2020ods,Murgia:2020ryi,Smith:2020rxx,Gomez-Valent:2021cbe}\footnote{See also \cite{Hill:2021yec,Poulin:2021bjr}, for the results obtained with the latest data release of the Atacama Cosmology Telescope \cite{ACT:2020gnv}.}. These discussions have been mainly based on the quantities $\sigma_8$ and $S_8$. As mentioned in Sec. \ref{sec:LCDM}, they are sometimes difficult to interpret as pure LSS estimators. Here we will also discuss the LSS in terms of $\sigma_{12}$ and $S_{12}$ \cite{Sanchez:2020vvb,Gomez-Valent:2021cbe} and will see that we can actually extract relevant information by analyzing these quantities.

\begin{table}[t!]
\centering
\begin{tabular}{|c ||c | c |}
 \multicolumn{1}{c}{EDE} & \multicolumn{2}{c}{Planck18+SNIa+BAO}
\\\hline
{\small Parameter} & {\small Marginalization}  & {\small PD}
\\\hline
$\omega_b$ & $0.02265^{+0.00019}_{-0.00017}$ & $0.02273^{+0.00021}_{-0.00020}$\\\hline
$\omega_{\rm cdm}$ & $0.1203^{+0.0039}_{-0.0016}$ & $0.1241^{+0.0027}_{-0.0025}$ \\\hline
$n_s$ & $0.971^{+0.009}_{-0.005}$ & $0.979\pm 0.008$ \\\hline
$\tau$ & $0.058\pm 0.007$ & $0.060\pm 0.007$ \\\hline
$\sigma_8$ & $0.816^{+0.011}_{-0.009}$ & $0.822^{+0.011}_{-0.010}$ \\\hline
$H_0$ [km/s/Mpc] & $68.4^{+1.3}_{-0.5}$ & $69.9^{+0.9}_{-1.0}$\\\hline
$f_{\rm ede}$ & $<0.048$ & $0.052^{+0.022}_{-0.021}$ \\\hline
$\theta_{\rm ini}$ & $>2.2$ & $>2.3$ \\\hline
$z_{\rm max}$ & $(3.9^{+4.2}_{-1.0})\cdot 10^{3}$ & $(4.2^{+3.0}_{-0.7})\cdot 10^{3}$ \\\hline\hline
$\log_{10}(m/{\rm eV})$ & $-26.66^{+0.66}_{-0.54}$ & $-27.24^{+0.38}_{-0.49}$ \\\hline
$\log_{10}(f/{\rm eV})$ & $26.56^{+0.44}_{-0.33}$ & $26.56^{+0.44}_{-0.40}$ \\\hline
$r_d$ [Mpc] & $146.9^{+0.5}_{-2.4}$ & $144.4\pm 1.5$ \\\hline
$M$ & $-19.394^{+0.033}_{-0.018}$ & $-19.352\pm 0.025$ \\\hline
$S_{8}$ & $0.825^{+0.012}_{-0.011}$ & $0.823^{+0.012}_{-0.009}$  \\\hline
$\sigma_{12}$ & $0.798^{+0.006}_{-0.007}$ & $0.797^{+0.006}_{-0.005}$ \\\hline
$S_{12}$ & $0.806^{+0.013}_{-0.009}$ & $0.812^{+0.012}_{-0.009}$ \\\hline
\end{tabular}
\label{tab:table}
\caption{Constraints on the parameters of the EDE model discussed in Sec. \ref{sec:EDE}. In the second column we show the values at the maximum of the marginalized one-dimensional posteriors and their $1\sigma$ uncertainties. In the third column we present the analogous quantities, extracted from the one-dimensional PDs. As cautioned in the main text, our PD method does not provide exact, but approximate results for the uncertainties, basically due to the noise in the MC chains. Our fast PD method is conceived to detect biases introduced by volume effects in the marginalization process. For a more accurate result one needs to carry out an exact PD analysis.}
\end{table}

As recently pointed out by the authors of \cite{Herold:2021ksg}, volume effects can have a sizable impact on the determination of the confidence intervals of the maximum fraction of EDE, $f_{\rm ede}$, allowed by the data. They found $f_{\rm ede}=0.072\pm 0.036$ at $68\%$ c.l. using the PD method an under the {\it Planck} 2018 and the BOSS DR12 full-shape likelihoods, whereas the authors of \cite{Ivanov:2020ril} found no peak at all for $f_{\rm ede}$ with the marginalization method, $f_{\rm ede}<0.072$ at $95\%$ c.l.

\begin{figure*}[t!]
\begin{center}               
\includegraphics[width=6in, height=7.0in]{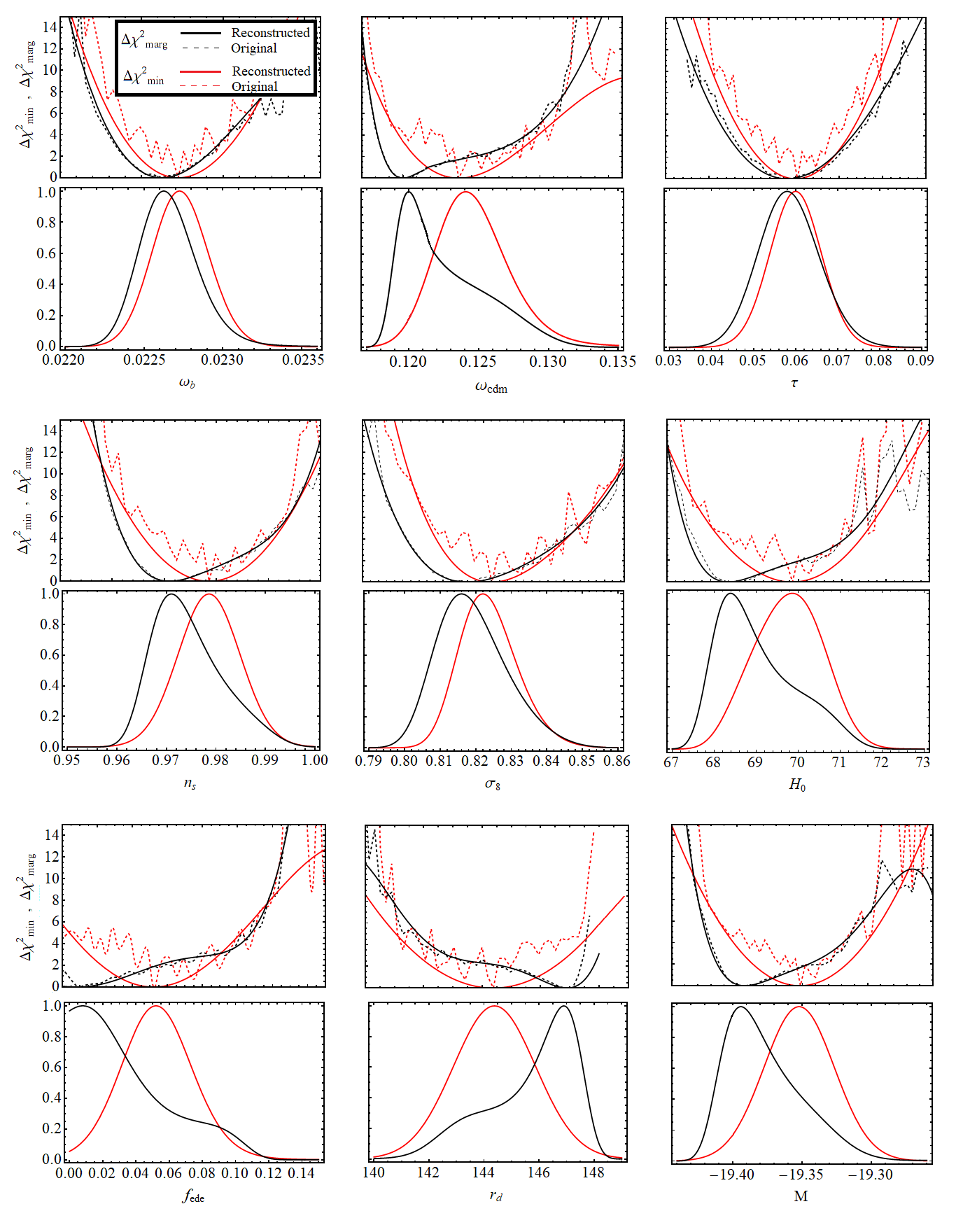}
\caption{Comparison between the results obtained with the marginalized posteriors (in black) and the PDs (in red) for several parameters listed in Table II for the EDE model. In the first, third and fifth rows of this figure we show the original and reconstructed profiles of $\Delta\chi^2$, whereas in the second, fourth and sixth rows we present the corresponding reconstructed distributions. Important ($\sim 1\sigma$) shifts are manifestly visible in most of these distributions. $H_0$ is expressed in km/s/Mpc, and $r_d$ in Mpc.}\label{fig:fig5}
\end{center}
\end{figure*}

\begin{figure}[t!]
\begin{center}
\includegraphics[width=3.3in, height=4in]{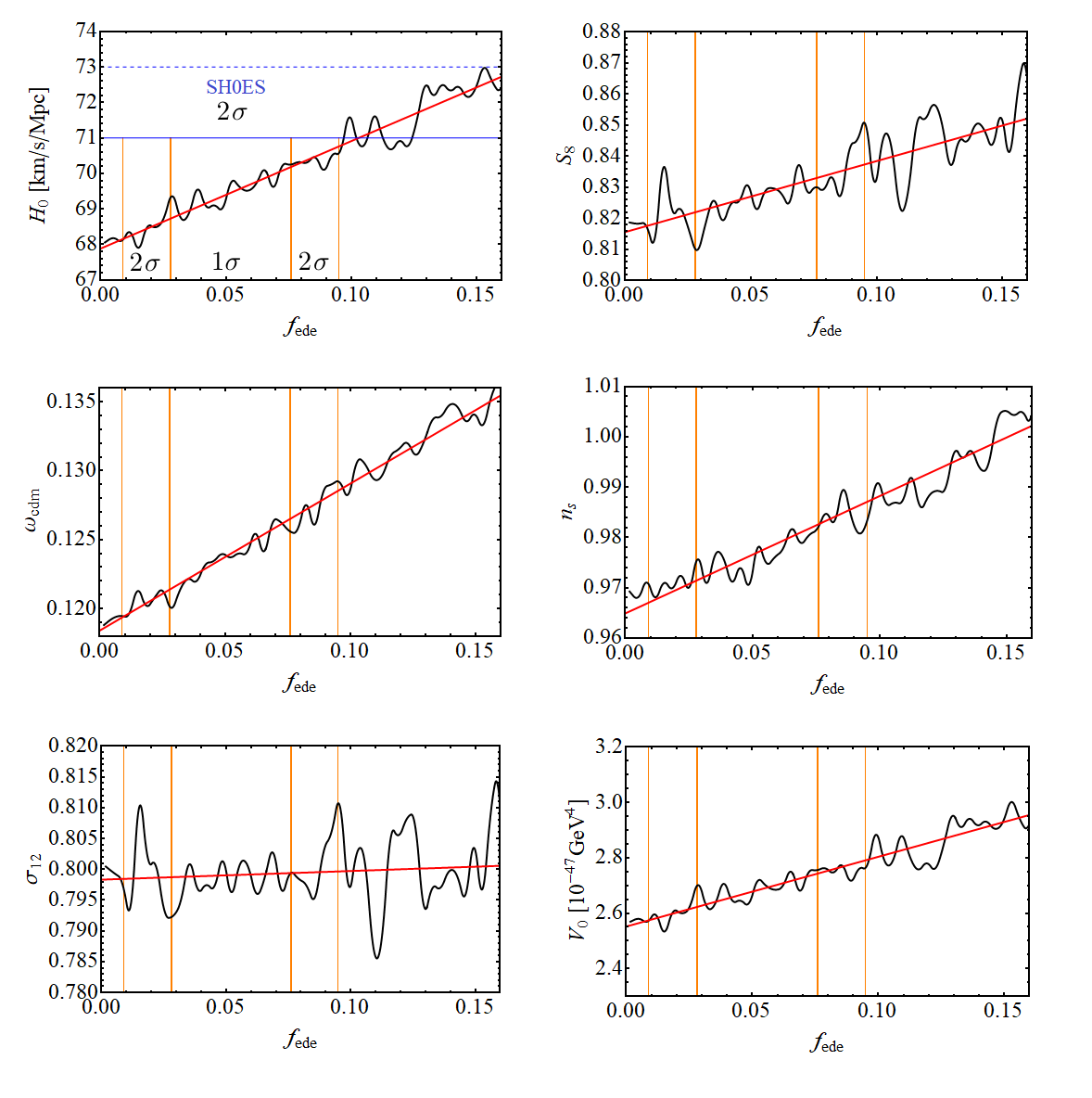}
\caption{The values of $(H_0,S_8,\omega_{\rm cdm},n_s,\sigma_{12},V_0)$ associated to $\chi^2_{\rm min}(f_{\rm ede})$, as a function of $f_{\rm ede}$ (in black). We also plot the straight lines obtained from the corresponding linear fits (in red), and show in all cases (in orange) the borders of the $1\sigma$ and $2\sigma$ bands of $f_{\rm ede}$ obtained from its PD. In the top left plot we include the central value (dashed line, in blue) and the border of the $2\sigma$ region (thick line, also in blue) of the last SH0ES measurement, $H_0=(73.04\pm 1.04)$ km/s/Mpc at $2\sigma$ c.l. \cite{Riess:2021jrx}.  See the comments in the main text, in Sec. \ref{sec:EDE}.}\label{fig:fig6}
\end{center}
\end{figure}

Here we reanalyze the EDE model using the Planck18+SNIa+BAO dataset, and report the constraints obtained directly from our MC Markov chains with the marginalization and PD methods, not only for $f_{\rm ede}$, but also for the other parameters of the theory. Previous works in the literature have only provided PD constraints on $f_{\rm ede}$ \cite{Herold:2021ksg}. Instead of using $(f,m)$ as input parameters we have implemented a shooting method to use $(f_{\rm ede},z_{\rm max})$, with the flat priors $f_{\rm ede}\in [0,1]$ and $z_{\rm max}\in [10^3,2\cdot 10^{4}]$. For a dedicated study on the impact that this choice has on the marginalized posteriors see \cite{Hill:2020osr}. For $\theta_{\rm ini}$ we also use a flat prior, $\theta_{\rm ini}\in [0,\pi]$. Our main results are presented in Table II and in Figs. \ref{fig:fig5} and \ref{fig:fig6}. In Table II we show the values of the parameters at the peaks of the one-dimensional marginalized posteriors and the PDs, together with their uncertainties. The marginalized value of the maximum EDE fraction reads, $f_{\rm ede}<0.048$ at $1\sigma$ c.l. Conversely, with the PD method we find a $\sim 2.5\sigma$ departure from the $\Lambda$CDM, $f_{\rm ede}=0.052^{+0.022}_{-0.021}$. The statistical significance of these results are very similar to those reported previously in \cite{Ivanov:2020ril,Herold:2021ksg}, although in our case we find the PD to peak at a somewhat smaller value. This parameter controls the departure of this EDE model from the standard model. When $f_{\rm ede}\sim 0$ we retrieve the $\Lambda$CDM, regardless of the values of $\theta_{\rm ini}$ and $z_{\rm max}$. Thus, there is a large volume fraction in parameter space that sticks to the concordance model, which still explains pretty well the data. This introduces volume effects, giving more weight to low values of $f_{\rm ede}$ in the marginalized posterior, and has also a direct impact on other parameters. The central value of $H_0$ obtained with the PD method is $1\sigma$ larger than the one obtained with the marginalization approach. They read $68.4^{+1.3}_{-0.5}$ (marg.) and $69.9^{+0.9}_{-1.0}$ km/s/Mpc (PD). This, in turn, is accompanied by important shifts in related quantities, since larger values of the Hubble parameter require lower values of $r_d$ and larger values of $M$ to ensure a good fit to the CMB+BAO and SNIa data, respectively. The parameters $r_d$ and $M$ still lie inside the range allowed by the low-redshift data, though \cite{Gomez-Valent:2021hda}. The aforesaid shifts can be easily appreciated in Fig. 5. Some parameters, as $\omega_b$ or $\tau$, do not seem to be extremely sensitive to the marginalization effects, but others, as those already mentioned before, exhibit important differences. This is also true for $\omega_{cdm}$, which has a peak at much larger values in the PD, simply because it is associated to the peak of $f_{\rm ede}$, which demands more dark matter to mitigate the increase of the early iSW effect. The uncertainties computed with our PD method are not exact, but the level of precision is sufficient to assess the impact of volume effects on our results. A more accurate analysis would require, of course, the exact computation of the PDs making use of efficient minimization routines, as in \cite{Herold:2021ksg}.

The Hubble tension is significantly alleviated when studied in terms of the PD. The value of $H_0$ lies $\sim 2.3\sigma$ below the last SH0ES measurement, $H_{0,\rm SH0ES}=(73.04\pm 1.04)$ km/s/Mpc \cite{Riess:2021jrx}. We can also quantify the tension by comparing the value of the absolute magnitude of SNIa obtained in our PD analysis, $M=-19.352\pm 0.025$, with the one measured by SH0ES in the first steps of the distance ladder, making use of Cepheids and other calibrators, $M_{\rm SH0ES}=-19.253\pm 0.027$ \cite{Riess:2021jrx}. The tension goes up to the $2.7\sigma$ c.l. in this case. The Hubble tension does not have in general the same statistical significance when assessed in terms of $M$ and $H_0$, as it was already pointed out in \cite{Gomez-Valent:2021cbe}. There are differences for the $\Lambda$CDM as well, where the Hubble tension reaches the $4.5\sigma$ ($H_0$) and $5.2\sigma$ ($M$) c.l. if we use the results obtained with the Planck18+SNIa+BAO dataset, cf. Table I. 

The distance ladder measurement from \cite{Freedman:2021ahq} obtained with the tip of the red giant branch (TRGB) method, $H_{0,{\rm TRGB}}=(69.8\pm 1.7)$ km/s/Mpc, has a central value that coincides almost exactly with the EDE PD result, but the statistical tension between the $\Lambda$CDM result for the Hubble parameter and $H_{0,{\rm TRGB}}$ is also derisory in this case ($<1\sigma$). 

The assessment of the statistical tension with the LSS data is considerably more subtle and tricky for a number of reasons that we will comment in this and subsequent paragraphs. First, we clearly observe that all the LSS estimators in ULA are slightly higher than in the $\Lambda$CDM, except $\sigma_{12}$ (cf. again Tables I and II). Let us try to understand why before rushing to conclusions. As explained before, the parameters $\sigma_8$ and $S_8$ are quite sensitive to $H_0$ because its value sets the scale that enters the window function in the computation of the {\it rms} of mass fluctuations \cite{Sanchez:2020vvb}. The fact that the typical values of the Hubble parameter are larger in this EDE model than in the $\Lambda$CDM contributes to increase the differences in these parameters obtained from the two models, being the ones in EDE larger. Conversely, $\sigma_{12}$ is computed at a fixed scale, and its values are quite similar in both models. This is telling us that EDE is actually able to keep similar amplitudes of the matter power spectrum to the $\Lambda$CDM while keeping also a good fit to the Planck18+SNIa+BAO data, even for relatively high values of $f_{\rm ede}\sim 0.05$. This does not contradict what we have explained previously. The model needs higher values of $\omega_{\rm cdm}$ to compensate the effects introduced by EDE in the pre-recombination epoch, of course, but in order to keep the good fit to the low-redshift data, e.g. SNIa, the model also increases the value of the cosmological constant $V_0$. This keeps the pressureless matter and DE fractions, $\Omega_m(a)$ and $\Omega_{\rm de}(a)=1-\Omega_m(a)$, at low redshifts close to the ones found in the $\Lambda$CDM, and leads to similar levels of the growth of matter perturbations in the late-time expansion, since the matter density contrast $\delta_m=\delta\rho_m/\rho_m$ is ruled by the following equation at subhorizon scales,

\begin{equation}\label{eq:dcEDE}
\delta_m^{\prime\prime}(a)+\frac{3}{2a}\delta_m^{\prime}(a)[2-\Omega_{m}(a)]-\frac{3}{2a^2}\Omega_m(a)\delta_m(a)=0\,.
\end{equation}   
which depends only on $\Omega_m(a)$, as in the standard model. The primes in Eq. \eqref{eq:dcEDE} denote derivatives with respect to the scale factor. The quantity $S_{12}$, instead, is sensitive to $\sigma_{12}$ and also to the current matter energy density, which is higher in EDE for the reason discussed before. This is why the value of $S_{12}$ is a bit larger than in the $\Lambda$CDM.

Here we do not analyze EDE under alternative datasets and, in particular, we do not study explicitly the impact of the LSS data. Nevertheless, we can make use of the results obtained with the Planck18+SNIa+BAO dataset to discuss some additional aspects related to the cosmological tensions. Fig. 6 will help us in this task. It shows the values of $(H_0,S_8,\omega_{\rm cdm},n_s,\sigma_{12},V_0)$ associated to $\chi^2_{\rm min}(f_{\rm ede})$, as a function of $f_{\rm ede}$. In other words, for each value of $f_{\rm ede}$, we look for the minimum $\chi^2$ in our Markov chain and save the parameters $(H_0,S_8,\omega_{\rm cdm},n_s,\sigma_{12},V_0)$ that lead to that minimum. Then we plot the dependence of each of these parameters with the maximum EDE fraction. The resulting curves are, again, a little bit noisy, since they are built directly from the chains, which have a limited number of points. In all cases, though, it is clear that they approximately follow a linear relationship. We have fitted a straight line, and plotted the result in red also in Fig. 6. There exists a positive correlation between all these parameters, although the one between $f_{\rm ede}$ and $\sigma_{12}$ is extremely small. The orange bands indicate the borders of the $1\sigma$ and $2\sigma$ regions obtained using the Planck18+SNIa+BAO dataset. The top left plot shows $H_0(f_{\rm ede})$. There we can appreciate in a more visual way the $2.3\sigma$ tension with the value of the Hubble parameter measured by SH0ES that we have discussed above. Notice also that reaching the SH0ES $1\sigma$ region ($\sim 72$ km/s/Mpc) requires $f_{\rm ede}$ to increase up to $\gtrsim 0.13$, which is certainly not favored by our dataset. This would force $\omega_{\rm cdm}\gtrsim 0.133$ and $n_s\gtrsim 0.995$, and values of $S_8\gtrsim 0.84$. The combined tomographic weak gravitational lensing analysis of the Kilo Degree Survey (KiDS +VIKING-450) and the Dark Energy Survey (DES-Y1), carried out under the assumption of the $\Lambda$CDM, led to the measurement $S_8=0.762^{+0.025}_{-0.024}$ \cite{Joudaki:2019pmv}. More recently, a joint analysis of the cross-correlations of galaxy positions and shears from DES-Y3 with CMB lensing maps from {\it Planck} and the South Pole Telescope (SPT) has given rise to a measurement with even a smaller central value, but with slightly larger uncertainties, $S_8=0.73^{+0.04}_{-0.03}$ \cite{DES:2022ign}. The EDE model under study does not seem to be capable of explaining these low values of $S_8$, while alleviating simultaneously the $H_0$ tension \cite{Hill:2020osr}. These tensions remain at best at $2-3\sigma$ c.l. At least, when the full {\it Planck} 2018 likelihood is considered. The addition of LSS data would make us obtain slightly lower values of $S_8$, but at the expense of worsening the Hubble tension. This conclusion was also reached in the tomographic study of \cite{Gomez-Valent:2021cbe}, using a more general fluid parametrization of a self-conserved EDE component with $c_s^2=1$. See also \cite{Moss:2021obd} for a similar study, in which the authors used a different binning approach, leaving also the EDE sound speed free in the MC; or \cite{Fondi:2022tfp} for the effect of curvature and a late-time dark energy dynamics. However, we should bear in mind that the aforementioned measurements of $S_8$ have been carried out assuming the $\Lambda$CDM, i.e. they are not model-independent. In the context of models with more parameters we expect the observational uncertainties to increase and the tension to be smaller. Actually, the plot $(f_{\rm ede},\sigma_{12})$ in Fig. \ref{fig:fig6} shows that the model is able to keep a reasonable amplitude of the linear matter power spectrum almost regardless of $f_{\rm ede}$. This is due to the reason explained in the preceding paragraph. The model is able to select larger values of the cosmological constant $V_0$ (see the last plot in Fig. \ref{fig:fig6}) and this allows to keep the energy fractions in the late-time universe close to the ones found in the standard model. According to these results, it seems that the model is able to decrease the Hubble tension without increasing the power of the matter fluctuations. It would be also very convenient to measure $S_{12}$ for the reasons discussed before, see again \cite{Sanchez:2020vvb}. This quantity is linked to a fixed scale and might allow us to better disentangle the tensions in $H_0$ and the LSS or, at least, to better understand up to what extent they are intertwined. 

If the TRGB measurement turned out to be closer to the true value of the Hubble parameter the $H_0$ tension would disappear in the context of this EDE model, and the $S_8$ one would be also small, below $2\sigma$. However, the $\Lambda$CDM would not perform much worse in this case. It would be still competitive.

The point that we want to remark the most in this paper is the impact of the marginalization effects on the results obtained from the MC chains, and their interpretation. Looking at the results of Table II and Fig. 5 it is clear that their impact is absolutely non-negligible in this EDE model. The marginalized posteriors basically underestimate the capability of EDE to explain the Planck18+SNIa+BAO data with larger EDE fractions, and this has an impact on the quantification of the cosmological tensions too. We agree with the authors of \cite{Herold:2021ksg}. It is important to employ the PD method to get unbiased information about the fitting capability of the model and evaluate properly the statistical significance of the tensions.


\subsubsection{Coupled dark energy}\label{sec:CDE}

\begin{figure*}[t!]
\begin{center}
\includegraphics[width=6.5in, height=2.75in]{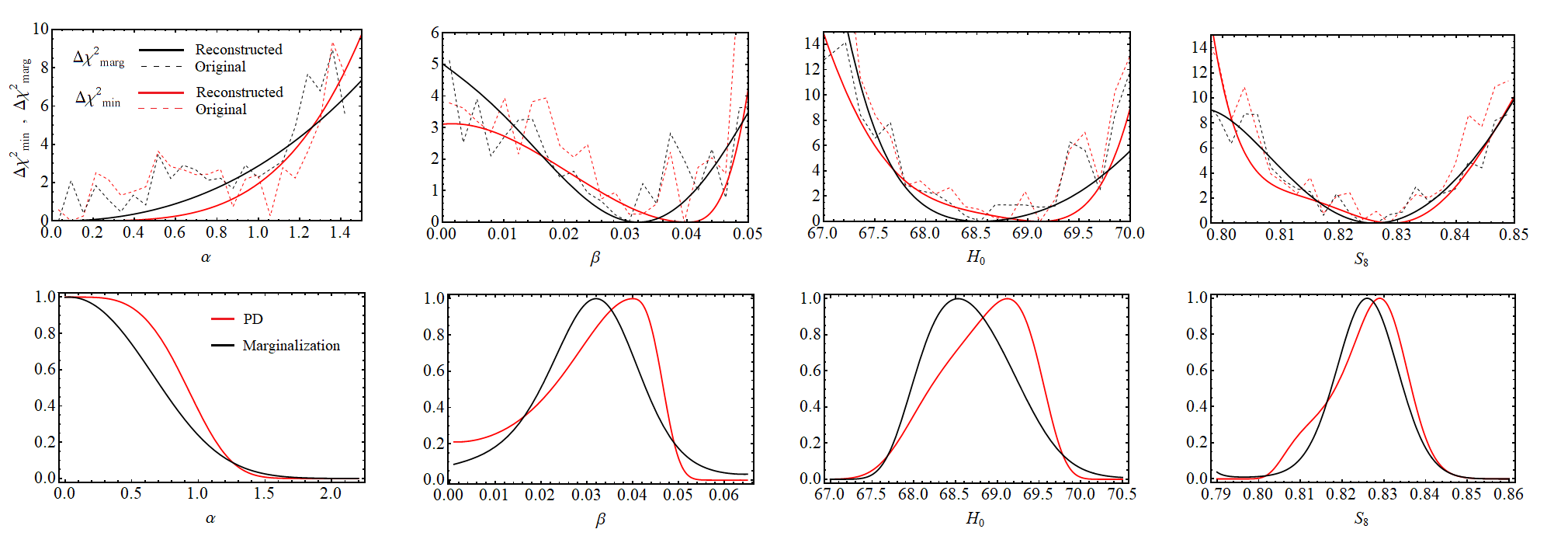}
\caption{{\it Upper plots:} The dashed curves are built directly from the MC Markov chains of the CDE model with the PD method (in red), $\Delta\chi^2_{\rm min}=-2\ln(\tilde{\mathcal{P}}/\tilde{\mathcal{P}}_{\rm max})$,  and the marginalization method (in black), $\Delta\chi^2_{\rm marg}=-2\ln(\mathcal{P}/\mathcal{P}_{\rm max})$, for the parameters $(\alpha,\beta,H_0,S_8)$. We also show the corresponding reconstructed functions using the same colors, but with solid lines. $H_0$ is expressed in km/s/Mpc; {\it Lower plots:} The smoothed distributions for the same parameters obtained with the PD and marginalization methods. See the comments in Sec. \ref{sec:CDE}.}\label{fig:fig7}
\end{center}
\end{figure*}  

We consider a CDE model with a constant coupling $\beta$ between the dark energy scalar field $\phi$ and cold dark matter \cite{Wetterich:1994bg,Amendola:1999er}, which modifies the conservation equations of the two species with respect to the ones in the uncoupled scenario. At the background level they read as follows, 

\begin{equation}
\ddot{\phi}+3H\dot{\phi}+\frac{\partial V}{\partial\phi}=\beta\kappa\rho_{\rm cdm}\,,
\end{equation}  
\begin{equation}\label{eq:densityCDM}
\dot{\rho}_{\rm cdm}+3H\rho_{\rm cdm}=-\beta\kappa\rho_{\rm cdm}\dot{\phi}\,,
\end{equation}                                               
with $\kappa=\sqrt{8\pi G}=\sqrt{8\pi}/M_{P}$ the inverse of the reduced Planck mass. $\beta$ is dimensionless. If we assume that the number of CDM particles is conserved throughout the expansion of the universe, equation \eqref{eq:densityCDM} leads to a decay of their mass,  

\begin{equation}
m_{\rm cdm}(z) = m_{\rm cdm}(z=0)e^{\beta\kappa[\phi(z=0)-\phi(z)]}\,.
\end{equation}
We use the Peebles-Ratra potential \cite{Peebles:1987ek,Ratra:1987rm}, 

\begin{equation}\label{eq:PR}
V(\phi)=V_0\phi^{-\alpha}\,,
\end{equation}
where $V_0$ is a constant with dimensions of energy to the $4+\alpha$ in natural units, and $\alpha>0$ a dimensionless constant\footnote{Notice that the constants $\alpha$ and $\beta$ used in this CDE model have nothing to do with the ones employed in the example of Sec. \ref{sec:SecI}. The constant $V_0$ is also different from the one used in Sec. \ref{sec:EDE}.}. This model has been discussed in detail both from the theoretical and phenomenological perspectives in a series of dedicated papers in the past, see e.g. \cite{Wetterich:1994bg,Amendola:1999er,Pettorino:2012ts,Pettorino:2013oxa,Planck:2015bue,Gomez-Valent:2020mqn}. For studies on CDE with alternative potentials, see also \cite{Xia:2013nua,vandeBruck:2016hpz,vandeBruck:2017idm,Agrawal:2019dlm,Gomez-Valent:2020mqn}. Here we summarize some of its basic features, and refer the reader to these references for more details.

\begin{table}[t!]
\centering
\begin{tabular}{|c ||c | c |}
 \multicolumn{1}{c}{CDE} & \multicolumn{2}{c}{Planck18+SNIa+BAO}
\\\hline
{\small Parameter} & {\small Marginalization}  & {\small PD}
\\\hline
$\omega_b$ & $0.02239^{+0.00008}_{-0.00010}$ & $0.02239^{+0.00009}_{-0.00011}$\\\hline
$\omega_{\rm cdm}$ & $0.1192\pm 0.0008$ & $0.1187^{+0.0011}_{-0.0006}$ \\\hline
$n_s$ & $0.967\pm 0.004$ & $0.967\pm 0.003$ \\\hline
$\tau$ & $0.056\pm 0.007$ & $0.060^{+0.005}_{-0.007}$ \\\hline
$\sigma_8$ & $0.826\pm 0.008$ & $0.831^{+0.004}_{-0.010}$ \\\hline
$H_0$ [km/s/Mpc] & $68.53^{+0.60}_{-0.46}$ & $69.12^{+0.37}_{-0.71}$\\\hline
$\alpha$ & $<0.58$ & $<0.62$ \\\hline
$\beta$ & $0.032^{+0.010}_{-0.012}$ & $0.040^{+0.006}_{-0.016}$ \\\hline\hline
$r_d$ [Mpc] & $146.71^{+0.20}_{-0.17}$ & $146.77^{+0.16}_{-0.22}$ \\\hline
$M$ & $-19.394^{+0.014}_{-0.012}$ & $-19.380^{+0.012}_{-0.019}$ \\\hline
$S_{8}$ & $0.826\pm 0.008$ & $0.829^{+0.007}_{-0.010}$  \\\hline
\end{tabular}
\label{tab:table}
\caption{As in Table II, but for the CDE model.}
\end{table}

During the radiation-dominated epoch the scalar field is almost frozen. The potential \eqref{eq:PR} is in charge of the late-time acceleration of the universe, so $V_0=\mathcal{O}(0.1M_P^2H_0^2)$ and, hence, many orders of magnitude lower than the radiation energy density during that era. In addition, $\alpha\ll 1$ in order the potential to behave close enough to a cosmological constant. Therefore, the scalar field potential does not play any role at large redshifts. On the other hand, the coupling does not become relevant until the matter-radiation equality time, when the CDM energy density starts to compete with the radiation one. Hence, for $z\gg z_{eq}$ $\phi$ remains constant in good approximation. At $z\sim z_{eq}$ the coupling enters into play and the mass of the CDM particles starts to decay in more or lesser extent depending on the value of $\beta$. This is regardless of its sign, so we can consider only $\beta>0$ in our MC runs, to simplify the analysis and speed up the convergence of the Markov chains. 

During the matter-dominated era, there is a scaling solution for the dark energy density that gives rise to a plateau in the dark energy fraction\footnote{The approximate symbol $\approx$ would become an equal symbol $=$ if baryons were also coupled to $\phi$ \cite{Amendola:1999er}.}, $\Omega_{\phi}\approx 2\beta^2/3$ \cite{Amendola:1999er}. It is absent when there is no interaction in the dark sector, i.e. when $\beta=0$. Due to the fifth force between the CDM particles, which is mediated by the scalar field, there is an enhancement of the matter energy density perturbations. This is clear from the equation that governs the evolution of the CDM density contrast $\delta_{\rm cdm}=\delta\rho_{\rm cdm}/\rho_{\rm cdm}$ at subhorizon scales,

\begin{eqnarray}\label{eq:DMcontrast}
\ddot{\delta}_{\rm cdm}+(2H-&&\beta\kappa\dot{\phi})\dot{\delta}_{\rm cdm}\phantom{XXXX}\\
&&-4\pi G[\rho_b\delta_b+\rho_{\rm cdm}\delta_{\rm cdm}(1+2\beta^2)]=0\,.\phantom{XX}\nonumber
\end{eqnarray}
The friction term decreases and the Poisson term increases when $\beta\ne 0$ because both, $\beta\dot{\phi}>0$ and $\beta^2>0$. Apart from this, for fixed current $\rho_{\rm cdm}$ the dark matter fluctuations start to grow sooner due to the increase of $m_{\rm cdm}$ in the past. All these facts make dark matter to cluster more efficiently. The aforementioned scaling solution corresponds to an unstable (saddle) fixed point, so the scaling regime is broken at some point, when the scalar field potential is of the same order of the matter energy density. At late times, the (mild) dynamics of $\phi$ is controlled mainly by the potential \eqref{eq:PR}, and dark energy behaves as quintessence. 

This model also allows to decrease the sound horizon at the baryon-drag epoch making use of the coupling to increase the mass $m_{\rm cdm}$ around the recombination time, what in turn (and in principle) can lead to larger values of $H_0$. However, by increasing $\beta$ we also get an enhancement of the matter power spectrum, which tends to worsen the tension with the LSS data, see e.g. \cite{Gomez-Valent:2020mqn}. This effect can be partially compensated by an increase of the late-time (quintessence) dynamics, but it limits the ability of the model to explain larger values of the Hubble parameter.

\begin{figure}[t!]
\begin{center}
\includegraphics[width=2.3in, height=3in]{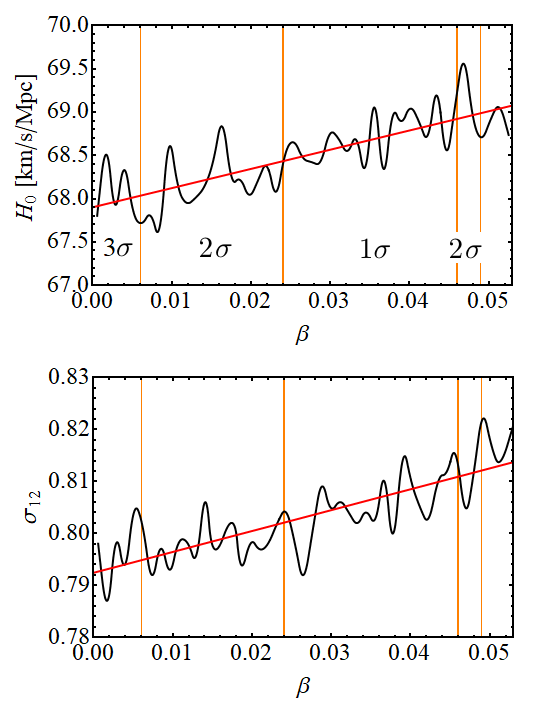}
\caption{The values of $(H_0,\sigma_{12})$ associated to $\chi^2_{\rm min}(\beta)$, in CDE, as a function of $\beta$ (in black). As in Fig. \ref{fig:fig6}, we also plot the linear fits (in red) and the borders of the $1\sigma$, $2\sigma$ and $3\sigma$ bands of $\beta$ obtained from the PDs (in orange).}\label{fig:CDEcorr}
\end{center}
\end{figure}

Here we reanalyze the fitting performance of this CDE model using the marginalization method with the Planck18+SNIa+BAO dataset, and compare the results with those obtained with the PD method to check whether volume effects have a sizable impact in this case. This has not been studied before in the literature. Apart from the $\Lambda$CDM parameters we also vary $\alpha$, $\beta$ and the initial conditions of the scalar field in our MC. It is important to remark that, in contrast to the EDE model studied in Sec. \ref{sec:EDE}, in this CDE model we have two parameters that control the deviations with respect to the standard model, $\alpha$ and $\beta$. If we set $\alpha=0$, dark matter and dark energy are coupled, but when dark matter becomes negligible due to the expansion of the universe the DE density tends to a constant, i.e. it has no dynamics at sufficiently large values of the cosmic time. If, instead, $\beta=0$, the CDM and DE are uncoupled and we recover the original Peebles-Ratra quintessence model \cite{Peebles:1987ek,Ratra:1987rm}. We retrieve the concordance model only in the joint limit $\alpha\to 0$ and $\beta\to 0$.

Our results are presented in Table III and Figs. \ref{fig:fig7} and \ref{fig:CDEcorr}. It is already evident from the numbers in Table III that the volume effects in this CDE model are not as important as in the EDE model studied in the previous section. The largest impact is observed on the parameters $\beta$ and $H_0$, which are of course positively correlated due to the reason discussed above (cf. the first plot in Fig. \ref{fig:CDEcorr}). The peak of the one-dimensional PD of the coupling is located at $\beta=0.040$, whereas the marginalized posterior peaks at $\beta=0.032$. However, Fig. 7 shows that the two distributions are quite similar. For example, the mean values computed from them read, $\beta=0.031\pm 0.011$ and $\beta=0.030^{+0.016}_{-0.006}$, respectively. In both cases there is a $\lesssim 3\sigma$ signal in favor of a non-null interaction in the dark sector, which comes mostly from the ability of the model to improve the description of the high-$l$ CMB temperature data with respect to the $\Lambda$CDM. Nevertheless, as mentioned before, these large values of $\beta$ increase the amount of LSS in the universe. At the best-fit point we find $S_8=0.829$ and $\sigma_{12}=0.814$, which are larger than the values found in the previous models. The Hubble parameter, instead, is $\sim 1$ km/s/Mpc larger than in the standard model,  according to the profile distribution. The Hubble tension with the SH0ES measurement \cite{Riess:2021jrx} remains, though, at $> 3\sigma$ c.l. This CDE model is not capable of alleviating significantly the cosmological tensions, and this is true even when the CMB lensing data from {\it Planck} is not included in the analysis \cite{Gomez-Valent:2020mqn}. The latter tends to favor lower values of $\alpha$ and larger values of $\beta$. Both facts help to increase the LSS in the universe. Other LSS datasets, as redshift-space distortions, tend to prefer larger values of $\alpha$ and lower values of $\beta$, to lower the amplitude of the matter power spectrum \cite{Gomez-Valent:2020mqn}. Contrary to the EDE model studied in Sec. \ref{sec:EDE}, CDE cannot produce larger values of $H_0$ (let us say, $H_0\gtrsim 70$ km/s/Mpc) by simultaneously keeping the amplitude of $P(k)$ at the level of the concordance model, see the lower plot of Fig. \ref{fig:CDEcorr}. The efficiency of the model is limited because the CDM mass at the last scattering surface is tightly constrained. Thus, the model cannot decrease significantly $r_{d}$ with respect to the values typically encountered in the $\Lambda$CDM, and the alleviation of the Hubble tension is therefore hindered. 

Regarding the parameter that controls the dynamics of the dark energy at late times, $\alpha$, we can see from both, Table III and Fig. \ref{fig:fig7}, that its value is compatible with $0$, since $\alpha\lesssim 0.6$ at $1\sigma$ c.l., and this result does not depend on the method we use to compute the confidence intervals.


\subsubsection{Brans-Dicke $\Lambda$CDM}\label{sec:BDLCDM}

Finally, we want to study the impact of volume effects also in the Brans-Dicke model \cite{BransDicke1961,dicke1962physical,brans1962mach} with a constant vacuum energy density, $\rho_\Lambda$ \cite{Avilez:2013dxa,deCruzPerez:2018cjx,SolaPeracaula:2019zsl,SolaPeracaula:2020vpg,Joudaki:2020shz,SolaPeracaula:2021gxi}. The latter was absent in the original BD theory, but it is a minimal ingredient needed to trigger the late-time acceleration of the universe. The resulting model was coined BD-$\Lambda$CDM in \cite{SolaPeracaula:2019zsl}, in contraposition to the standard GR-$\Lambda$CDM studied in Sec. \ref{sec:LCDM}, which is built on the basis of General Relativity (GR). The action of this scalar-tensor theory of gravity reads, $S_{\rm BD}=\tilde{S}_{\rm BD}+S_m$, with

\begin{equation}\label{eq:BDaction}
\tilde{S}_{\rm BD}=\int d^{4}x\sqrt{-g}\left[\frac{1}{16\pi}\left(R\psi-\frac{\omega_{\rm BD}}{\psi}\partial^{\mu}\psi\partial_{\mu}\psi\right)-\rho_\Lambda\right]\,,
\end{equation}
and $S_m$ the action for the matter fields, which are not directly coupled to the BD scalar $\psi$. The BD field has dimensions of energy squared in natural units, and can be thought of as the inverse of the effective gravitational coupling, which in this model is dynamical. The modified Klein-Gordon equation that governs its evolution takes the following form in a flat FLRW spacetime, 

\begin{equation}\label{eq:KGpsi}
\ddot{\psi}+3H\dot{\psi}=\frac{8\pi}{2\omega_{\rm BD}+3}(\rho-3p)\,,
\end{equation}
where $\rho$ and $p$ are the total energy density and pressure in the universe, respectively. The BD parameter $\omega_{\rm BD}$ controls the dynamics of the scalar field. In the limit $\omega_{\rm BD}\to \infty$ the field $\psi$ has only a decaying mode. When it goes to zero the field tends to a constant, and if it is equal to $G^{-1}_N$, with $G_N$ Newton's constant, the model reduces to the standard GR-$\Lambda$CDM. Cosmological observations already force $\omega_{\rm BD}\gtrsim 300$ (see e.g. \cite{SolaPeracaula:2019zsl,SolaPeracaula:2020vpg,SolaPeracaula:2021gxi}), so it is useful to define the (small) quantity 

\begin{equation}\label{eq:epsilonBD}
\epsilon_{\rm BD}=\frac{1}{\omega_{\rm BD}}\ll1\,,
\end{equation}  
which parametrizes the slow dynamics of $\psi$. We will provide our constraints on \eqref{eq:epsilonBD}, instead of $\omega_{\rm BD}$. 

The various matter species filling the universe are self-conserved, as in the concordance model, and the modified Friedmann and pressure equations read, respectively, 

\begin{equation}
3H^2 + 3H\frac{\dot{\psi}}{\psi} -\frac{\omega_{\rm BD}}{2}\left(\frac{\dot{\psi}}{\psi}\right)^2 = \frac{8\pi}{\psi}\rho\label{eq:Friedmannequation}
\end{equation}
and
\begin{equation}
2\dot{H} + 3H^2 + \frac{\ddot{\psi}}{\psi} + 2H\frac{\dot{\psi}}{\psi} + \frac{\omega_{\rm BD}}{2}\left(\frac{\dot{\psi}}{\psi}\right)^2 = -\frac{8\pi}{\psi}p\,.\label{eq:pressureequation}
\end{equation}
The scalar field remains basically constant during the radiation-dominated era, since $\rho-3p=\rho_m\ll \rho_r$ in that period of the cosmic expansion, so the {\it rhs} of equation \eqref{eq:KGpsi} is completely negligible. Once the scalar field reaches the constant solution it has no evolution until the matter-radiation equality time, when the field acquires some dynamics and starts to change with the expansion. If $\epsilon_{\rm BD}>0$ the scalar field grows, i.e. the effective gravitational coupling at cosmological scales decreases. Conversely, if $\epsilon_{\rm BD}<0$ the field $\psi$ decreases and the effective $G$ grows with the cosmic time. For $|\epsilon_{\rm BD}|\ll 1$ it evolves logarithmically with the scale factor, cf. \cite{SolaPeracaula:2020vpg}.

\begin{figure*}[t!]
\begin{center}
\includegraphics[width=6.5in, height=2.75in]{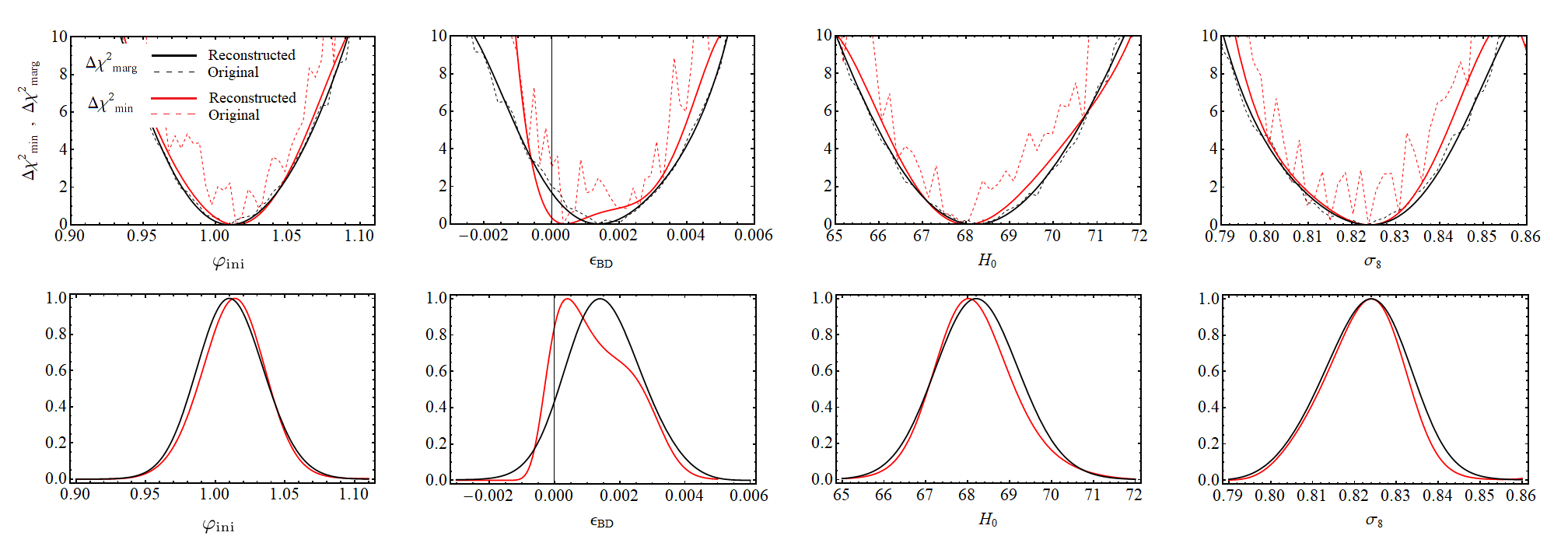}
\caption{As in Fig. \ref{fig:fig7}, but for the parameters $(\varphi_{\rm ini},\epsilon_{\rm BD},H_0,\sigma_8)$ of the BD-$\Lambda$CDM. $H_0$ is expressed in km/s/Mpc. See the comments in Sec. \ref{sec:BDLCDM}.}\label{fig:fig8}
\end{center}
\end{figure*}

Matter perturbations at deep subhorizon scales are ruled by the following equation for the total matter density contrast, 

\begin{equation}\label{eq:ExactPerturScaleFactor}
\ddot{\delta}_m+2H\dot{\delta}_m-\frac{4\pi}{\psi}\rho_m\delta_m\left(\frac{2+4\epsilon_{\rm BD}}{2+3\epsilon_{\rm BD}}\right)=0\,.
\end{equation}
Notice that the effective $G=1/\psi$ that enters the background expressions, e.g. the Friedmann equation \eqref{eq:Friedmannequation}, is not the same that controls the gravitational force between two test masses and the clustering at large scales, which reads,

\begin{equation}
G=\frac{1}{\psi}\left(\frac{2+4\omega_{\rm BD}}{2+3\omega_{\rm BD}}\right)\,,
\end{equation}
cf. \cite{SolaPeracaula:2020vpg} for details. During matter domination, $\delta_m\sim a^{1+\epsilon_{\rm BD}}$. Thus, negative values of the parameter $\epsilon_{\rm BD}$ make matter to cluster less efficiently than in the standard model. On the other hand, a larger value of the effective $G$ at cosmological scales can give rise to larger values of the Hubble parameter both, at early and late times, and this can alleviate the Hubble tension. Overall, this is also possible thanks to an increase of the spectral index $n_s$, see \cite{SolaPeracaula:2020vpg}. 

\begin{table}[t!]
\centering
\begin{tabular}{|c ||c | c |}
 \multicolumn{1}{c}{BD-$\Lambda$CDM} & \multicolumn{2}{c}{Planck18+SNIa+BAO}
\\\hline
{\small Parameter} & {\small Marginalization}  & {\small PD} 
\\\hline
$\omega_b$ & $0.02235\pm 0.00020$ & $0.02228^{+0.00023}_{-0.00014}$ \\\hline
$\omega_{\rm cdm}$ & $0.1200^{+0.0013}_{-0.0011}$ & $0.1207^{+0.0009}_{-0.0012}$  \\\hline
$n_s$ & $0.964\pm 0.007$ & $0.963\pm 0.007$  \\\hline
$\tau$ & $0.055^{+0.008}_{-0.007}$ & $0.058^{+0.006}_{-0.007}$  \\\hline
$\sigma_8$ & $0.824^{+0.010}_{-0.011}$ & $0.824^{+0.008}_{-0.010}$  \\\hline
$H_0$ [km/s/Mpc] & $68.20^{+1.03}_{-1.00}$ & $68.02^{+1.00}_{-0.88}$ \\\hline
$\epsilon_{\rm BD}$ & $0.0014^{+0.0012}_{-0.0011}$ & $0.0005^{+0.0017}_{-0.0006}$  \\\hline
$\varphi_{\rm ini}$ & $1.010^{+0.025}_{-0.024}$ & $1.014^{+0.022}_{-0.024}$   \\\hline\hline
$r_d$ [Mpc] & $147.88^{+2.00}_{-1.93}$ & $148.22^{+1.67}_{-1.91}$ \\\hline
$M$ & $-19.411^{+0.034}_{-0.030}$ & $-19.441^{+0.060}_{-0.014}$ \\\hline
$\varphi(z=0)$ & $1.020^{+0.031}_{-0.027}$ & $1.026^{+0.026}_{-0.030}$  \\\hline
$\dot{G}/G$ [$10^{-13}$ yr$^{-1}$] & $-1.61^{+1.31}_{-1.25}$ & $-1.61^{+1.52}_{-1.03}$  \\\hline
$\sigma_{12}$ & $0.809^{+0.011}_{-0.010}$ & $0.812^{+0.011}_{-0.010}$  \\\hline
$S_{12}$ & $0.818^{+0.012}_{-0.014}$ & $0.820^{+0.010}_{-0.014}$ \\\hline
\end{tabular}
\label{tab:table}
\caption{As in Table II and III, but for the BD-$\Lambda$CDM model.}
\end{table}

The BD-$\Lambda$CDM has been confronted with observations multiple times in the last years \cite{Avilez:2013dxa,deCruzPerez:2018cjx,SolaPeracaula:2019zsl,SolaPeracaula:2020vpg,Joudaki:2020shz,SolaPeracaula:2021gxi}. For instance, in the most recent work \cite{SolaPeracaula:2021gxi} the authors showed that in the absence of the {\it Planck} 2018 high-$l$ CMB polarization and lensing data, but still under a very rich dataset including the {\it Planck} 2018 full temperature and low-$l$ polarization likelihoods, together with the state-of-the-art data on SNIa, BAO, cosmic chronometers and redshift-space distortions, it is possible to loosen the $H_0$ tension in the BD-$\Lambda$CDM, while keeping $\sigma_8\sim 0.790$. They found $H_0=(69.30^{+1.38}_{-1.35})$ km/s/Mpc without including the prior on $H_0$ from SH0ES \cite{Reid:2019tiq} in the MC analysis, and $H_0=(71.23^{+1.01}_{-1.02})$ km/s/Mpc considering it. The $H_0$ tension is lowered in these cases to $\sim 2.2\sigma$ and $\sim 1.2\sigma$ c.l., respectively. We refer the reader to \cite{SolaPeracaula:2020vpg} for a detailed discussion on theoretical aspects of the BD-$\Lambda$CDM model, the cosmological tensions, and fitting results obtained under a wide variety of datasets.

As mentioned before, the model typically needs negative values of $\epsilon_{\rm BD}$ to keep under control the $\sigma_8$ tension (with $|\epsilon_{\rm BD}|\sim 10^{-3}$). Values of the effective $G$ $\sim 4-9\%$ larger than $G_N$ are still allowed by the data, especially when the CMB lensing and high-$l$ polarization data are not included. This allows to loosen the Hubble tension \cite{SolaPeracaula:2020vpg}. When the SH0ES prior is included in the analysis it is even found a departure of the cosmological $G$ with respect to $G_N$ at $\sim 2.7\sigma$ c.l. \cite{SolaPeracaula:2021gxi}. However, the modified gravity effects should be screened in the Solar System to respect the local constraints that force the local values of $\epsilon_{\rm BD}$ and $G$ to satisfy $|\epsilon_{\rm BD}|\lesssim 10^{-5}$ \cite{Bertotti:2003rm} and $G\approx G_N$. Suppressing the dynamics of the scalar field is relatively easy adding higher-derivative terms to the original action \eqref{eq:BDaction}, e.g. using Vainshtein or K-mouflage screening mechanisms \cite{Gomez-Valent:2021joz}. Explaining $\gtrsim \mathcal{O}(1\%)$ differences between the gravitational coupling at cosmological and local scales might not be so straightforward, see \cite{Gomez-Valent:2021joz}. Here, though, we take for granted the existence of an efficient screening mechanism capable of retrieving standard GR in our vicinity\footnote{Another possibility would be to consider a very late-time transition of $G$, happening at $z\lesssim 0.01$ \cite{Marra:2021fvf}, and leading to $\psi(z=0)=1/G_N$ at present \cite{Gomez-Valent:2021joz}. In this case the Hubble tension would be automatically solved, although one would need to explain still the cause of such an abrupt transition.}, as it was done in the previous works in the literature, and use the same Planck18+SNIa+BAO dataset employed before to constrain the other models. 

\begin{figure}[t!]
\begin{center}
\includegraphics[width=3.2in, height=2.7in]{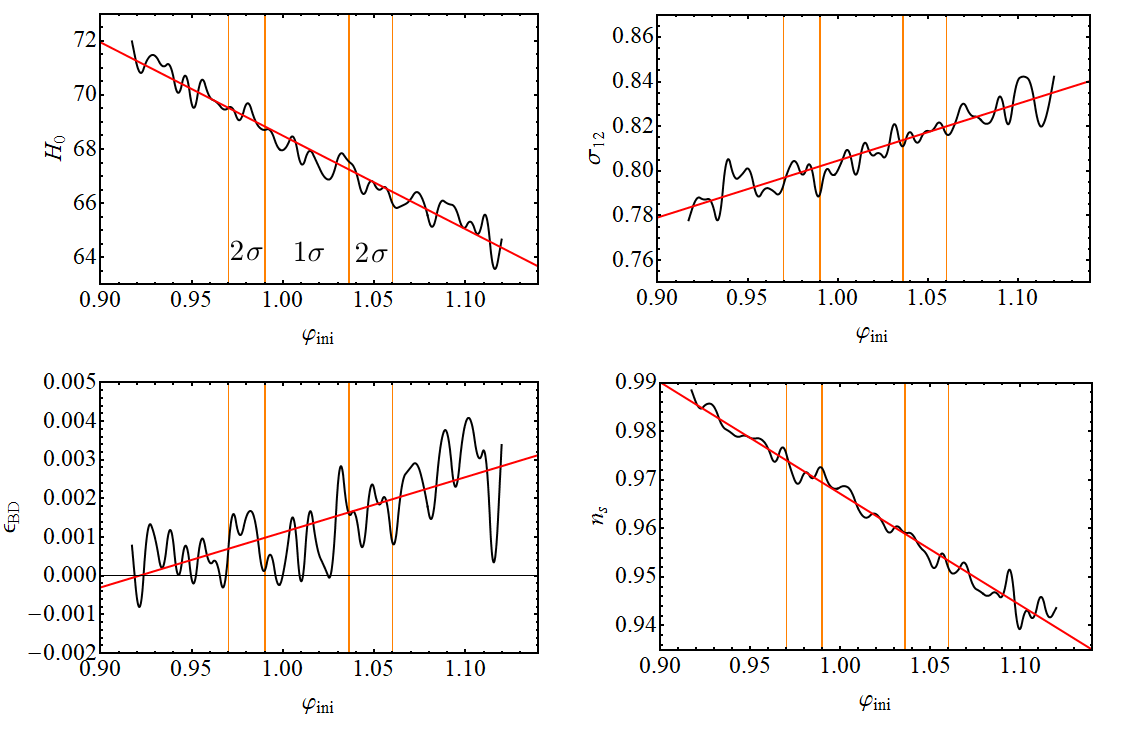}
\caption{The values of $(H_0,\sigma_{12},\epsilon_{\rm BD},n_{s})$ associated to $\chi^2_{\rm min}(\varphi_{\rm ini})$, in the BD-$\Lambda$CDM, as a function of $\varphi_{\rm ini}$ (in black). We plot the linear fits (in red) and the borders of the $1\sigma$ and $2\sigma$ of $\varphi_{\rm ini}$ obtained from the PDs (in orange).}\label{fig:corr_BDLCDM}
\end{center}
\end{figure}

The parameter space of the BD-$\Lambda$CDM model is composed by the six GR-$\Lambda$CDM parameters (cf. Sec. \ref{sec:LCDM}), together with $\epsilon_{\rm BD}$ and the initial value of the scalar field, which we express in terms of the dimensionless quantity $\varphi_{\rm ini}\equiv G_N\psi_{\rm ini}$. The current value of the normalized scalar field $\varphi(z=0)=G_N\psi(z=0)$ is obtained as a derived parameter, and also the relative variation of the effective $G$ with the cosmic time evaluated at present, $\dot{G}/G$. Our main results are presented in Table IV and Figs. \ref{fig:fig8} and \ref{fig:corr_BDLCDM}. Most of the parameters are not sensitive to the method employed to extract their constraints. For some of them, as $\epsilon_{\rm BD}$, the peak obtained with the PD method is shifted with respect to the one obtained with the marginalization approach. However, these shifts are not statistically significant. We can conclude that volume effects do not introduce an important bias in the analysis of the BD-$\Lambda$CDM. 

\begin{figure*}[t!]
\begin{center}
\includegraphics[width=6.2in, height=2.2in]{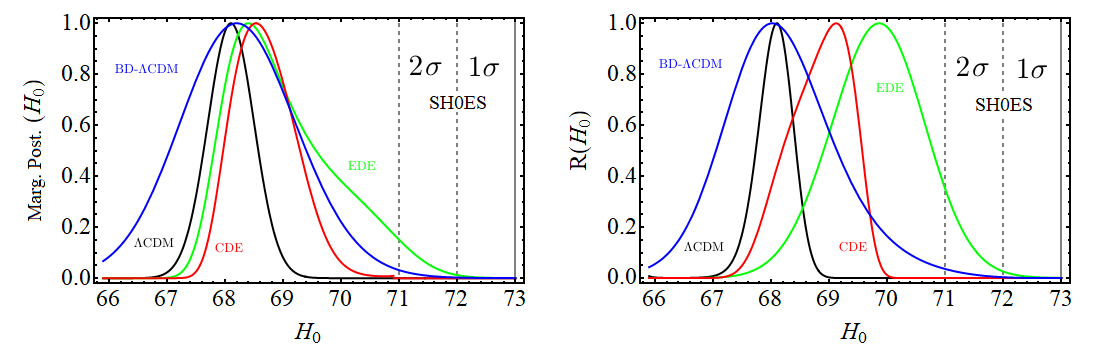}
\caption{Comparison of the marginalized posteriors (left plot) and PDs (right plot) of $H_0$ (in km/s/Mpc) for the four models studied in this paper. We also show the $1\sigma$ and $2\sigma$ vertical bands of the SH0ES measurement \cite{Riess:2021jrx}.}\label{fig:fig9}
\end{center}
\end{figure*}

The model is able to produce considerably larger values of $H_0$ than the concordance model. This can be easily appreciated in Figs. \ref{fig:fig8} and \ref{fig:fig9}. The model seems also able to explain larger values than CDE, in the right tail of the distribution, although performs worse than EDE concerning the Hubble tension, considering of course the Planck18+SNIa+BAO dataset. There is still a $\sim 2.5\sigma$ tension with SH0ES. The model has, though, some interesting features that we want to remark. The parameter $\varphi_{\rm ini}$ is positively correlated with $\sigma_{12}$ and negatively correlated with $H_0$, see Fig. \ref{fig:corr_BDLCDM}. This helps to fight simultaneously against the two tensions, with values of $\varphi_{\rm ini}<1$, especially when we remove the CMB high-$l$ polarization and CMB lensing data from {\it Planck} 2018 \cite{SolaPeracaula:2019zsl,SolaPeracaula:2020vpg,SolaPeracaula:2021gxi}. The latter limit a lot the minimum value of $\varphi_{\rm ini}$. These correlations are very interesting and are a feature that is not shared by the other models. The correlation coefficient between $\beta$ and the Hubble parameter is positive in CDE, but the one between $\beta$ and $\sigma_{12}$ is also positive, see Fig. \ref{fig:CDEcorr}. Hence, $\beta$ cannot help to alleviate both tensions at a time. In EDE, instead, $H_0$ grows with $f_{\rm ede}$, but $\sigma_{12}$ cannot be significantly decreased simultaneously because the correlation coefficient between $f_{\rm ede}$ and $\sigma_{12}$ is very small. $\sigma_{12}$ can be kept, though, to similar values to the $\Lambda$CDM, cf. Fig. \ref{fig:fig6}.

If the CMB lensing and polarization data from {\it Planck} are subject to unaccounted systematics, the BD-$\Lambda$CDM cosmology could be an interesting option to alleviate the tensions \cite{SolaPeracaula:2019zsl,SolaPeracaula:2020vpg,SolaPeracaula:2021gxi}. If these datasets are not affected by systematics, it seems that EDE can perform better to explain the SH0ES measurement, although it is unable to produce a lower amplitude of the matter power spectrum. The exact phenomenological status of these models can vary depending on the dataset under consideration. It will be important to track in the future the new data and continue testing their performance. Here we have seen that marginalization effects can be relevant in some cases, also in the discussion of the cosmological tensions, and have explored a method that allows us to assess their impact in a fast way evaluating the profile distributions directly from the MC Markov chains.


\section{Conclusions}\label{sec:conclusions}

In this paper we have motivated the use of the one-dimensional profile distributions (PDs) as a means to compute the constraints on the model parameters in Bayesian analyses. We are forced to use Monte Carlo routines to sample multivariate distributions in high-dimensional parameter spaces. The marginalization procedure usually employed to obtain the confidence intervals of the parameters can induce a bias in the interpretation of the results that are encoded in the Markov chains, which cannot be quantified by just looking at the marginalized results. The profile distributions allow us to get rid of the volume effects, but their computation is expensive in terms of computational time, and this can be a problem when we are interested in many parameters. Preliminary Monte Carlo runs are needed to estimate the regions in parameter space where we expect to find the bulk of the probability and the best-fit point, to improve the efficiency of the PD method. We have shown that we can take advantage of these preliminary MC runs to estimate the PDs directly from the Markov chains. This can be done very fast, and inform us about the existence of biases introduced in the marginalization step, and also about the need to perform a more exhaustive (exact) PD analysis. Although the results obtained with the profile distributions built from the Markov chains are subject to some degree of noise, they already let us identify for which parameters the volume effects are important, without carrying out the full (exact) minimization procedure. We remark, though, that the latter is of course needed for a more precise estimation of the PDs and the uncertainties of the parameters that are derived from them. In particular, we point out that our method might lead in general to larger relative errors in the estimation of the minimum $\chi^2$ in the tails of the distributions, since the Monte Carlo does not spend enough time in these regions of parameter space. This is a problem that can be mitigated, though, by obtaining longer chains, whereas the marginalization issues persist regardless of their length. Our method can be applied to the main parameters that are varied in the Monte Carlo, but not only so. We can also apply it to the nuisance and derived parameters of interest, with no significant additional effort. 

We have illustrated the potential impact of volume effects through a very pedagogical example in two dimensions, and how to solve the problem with the use of the PD method. Then, we have applied it to the analysis of four different cosmological models, namely, the standard $\Lambda$CDM, early dark energy (EDE), coupled dark energy (CDE), and Brans-Dicke cosmology with a constant vacuum energy density (BD-$\Lambda$CDM). We have used the full {\it Planck} 2018 likelihood together with the Pantheon compilation of supernovae of Type Ia and data on baryon acoustic oscillations to constrain these models. We have shown that the concordance model is basically unaffected by marginalization issues, as reported previously by the {\it Planck} 2013 collaboration \cite{Planck:2013nga}. For EDE, instead, they have a huge impact. We find a similar result to \cite{Herold:2021ksg} for the maximum EDE fraction. With the PD method we find $f_{\rm ede}=0.052^{+0.022}_{-0.021}$ at $1\sigma$ c.l., which is $\sim 2.5\sigma$ away from the $\Lambda$CDM, i.e. from $f_{\rm ede}=0$. This is in contrast to the marginalized result, which clearly hides this peak, giving $f_{\rm ede}<0.048$ at $1\sigma$ c.l. We have also studied for the first time in the literature the impact of this issue on the other parameters of the theory, revealing important shifts in the distributions of $\omega_{\rm cdm}$ and the parameters $(H_0,r_d,M)$, which are of utmost importance for the discussion on the Hubble tension. We have studied the status of the cosmological tensions in the context of the Planck18+SNIa+BAO data, and have pointed out that it is possible to keep the parameter $\sigma_{12}$ in this EDE model close to the value found in the $\Lambda$CDM even for relatively high values of $f_{\rm ede}$. This is because the model leaves room to the cosmological constant to increase in order to stabilize the matter and dark energy fractions in the late-time universe. Finally, we have also studied the impact of marginalization on CDE and the BD-$\Lambda$CDM. Despite being non-zero, the impact of volume effects is lower in these cases. 

The PD approach can be also applied to produce two-dimensional confidence regions. This is useful to study e.g. correlations and tensions involving two parameters. We have not presented them in this paper because the results were too noisy and were difficult to smooth. One possible way to solve this issue is by adding more points to the Markov chains. This would slow the process, but if non-Gaussianities are not excessively large one can use methods that allow to boost the Monte Carlo exploration of the parameter space, e.g. substituting the exact likelihood by a non-Gaussian fit, see \cite{Amendola:2020qkb,Rizzato:2022hbu}. The alternative is to perform the exact computation of the two-dimensional PDs, using e.g. \texttt{MontePython} in combination with some minimization code like Py-BOBYQA \cite{BOBYQA,PyBOBYQA}. This is left for a future work. The results obtained in the MC, despite being noisy, can also help to speed the computation of the exact PDs.

The main conclusion of this work is that volume effects can play a role in some cosmological models, and under some datasets. EDE is a clear example. It is important to have a good control on all these effects to better assess the fitting performance of these models and the statistical significance of the cosmological tensions. We strongly recommend the use of the PD method to detect potential marginalization biases in MC studies. Our method offers the possibility of doing so in an efficient way.  


\vspace{0.25cm}
\noindent {\bf Acknowledgements}
\newline
\newline
\noindent The author is funded by the Istituto Nazionale di Fisica Nucleare (INFN) through the project of the InDark INFN Special Initiative: ``Dark Energy and Modified Gravity Models in the light of Low-Redshift Observations'' (n. 22425/2020). He is grateful to the Institute for Theoretical Physics (ITP) Heidelberg for letting him use its computational facilities remotely. He also wants to thank his collaborators for previous and ongoing work on the models discussed in this paper: Prof. Luca Amendola, Prof. Joan Sol\`a Peracaula, Prof. Christof Wetterich, Dr. Javier de Cruz P\'erez, Dr. Valeria Pettorino, Lisa Goh, Prajwal Hassan Puttasidappa, Cristian Moreno-Pulino and Ziyang Zheng. Finally, he thanks Dr. Javier Carrón for discussions on the topic dealt with in this paper, and the Referee for their very insightful comments and suggestions, which have certainly helped the author to improve the presentation of this work.


\bibliographystyle{apsrev4-1}
\bibliography{Marginalization}

\begin{thebibliography}{78}%
\makeatletter
\providecommand \@ifxundefined [1]{%
 \@ifx{#1\undefined}
}%
\providecommand \@ifnum [1]{%
 \ifnum #1\expandafter \@firstoftwo
 \else \expandafter \@secondoftwo
 \fi
}%
\providecommand \@ifx [1]{%
 \ifx #1\expandafter \@firstoftwo
 \else \expandafter \@secondoftwo
 \fi
}%
\providecommand \natexlab [1]{#1}%
\providecommand \enquote  [1]{``#1''}%
\providecommand \bibnamefont  [1]{#1}%
\providecommand \bibfnamefont [1]{#1}%
\providecommand \citenamefont [1]{#1}%
\providecommand \href@noop [0]{\@secondoftwo}%
\providecommand \href [0]{\begingroup \@sanitize@url \@href}%
\providecommand \@href[1]{\@@startlink{#1}\@@href}%
\providecommand \@@href[1]{\endgroup#1\@@endlink}%
\providecommand \@sanitize@url [0]{\catcode `\\12\catcode `\$12\catcode
  `\&12\catcode `\#12\catcode `\^12\catcode `\_12\catcode `\%12\relax}%
\providecommand \@@startlink[1]{}%
\providecommand \@@endlink[0]{}%
\providecommand \url  [0]{\begingroup\@sanitize@url \@url }%
\providecommand \@url [1]{\endgroup\@href {#1}{\urlprefix }}%
\providecommand \urlprefix  [0]{URL }%
\providecommand \Eprint [0]{\href }%
\providecommand \doibase [0]{http://dx.doi.org/}%
\providecommand \selectlanguage [0]{\@gobble}%
\providecommand \bibinfo  [0]{\@secondoftwo}%
\providecommand \bibfield  [0]{\@secondoftwo}%
\providecommand \translation [1]{[#1]}%
\providecommand \BibitemOpen [0]{}%
\providecommand \bibitemStop [0]{}%
\providecommand \bibitemNoStop [0]{.\EOS\space}%
\providecommand \EOS [0]{\spacefactor3000\relax}%
\providecommand \BibitemShut  [1]{\csname bibitem#1\endcsname}%
\let\auto@bib@innerbib\@empty
\bibitem [{\citenamefont {Trotta}(2017)}]{Trotta:2017wnx}%
  \BibitemOpen
  \bibfield  {author} {\bibinfo {author} {\bibfnamefont {R.}~\bibnamefont
  {Trotta}}\ }(\bibinfo {year} {2017})\ \Eprint
  {http://arxiv.org/abs/1701.01467} {arXiv:1701.01467 [astro-ph.CO]}
  \BibitemShut {NoStop}%
\bibitem [{\citenamefont {Kamionkowski}\ \emph {et~al.}(2014)\citenamefont
  {Kamionkowski}, \citenamefont {Pradler},\ and\ \citenamefont
  {Walker}}]{Kamionkowski:2014zda}%
  \BibitemOpen
  \bibfield  {author} {\bibinfo {author} {\bibfnamefont {M.}~\bibnamefont
  {Kamionkowski}}, \bibinfo {author} {\bibfnamefont {J.}~\bibnamefont
  {Pradler}}, \ and\ \bibinfo {author} {\bibfnamefont {D.~G.~E.}\ \bibnamefont
  {Walker}},\ }\href {\doibase 10.1103/PhysRevLett.113.251302} {\bibfield
  {journal} {\bibinfo  {journal} {Phys. Rev. Lett.}\ }\textbf {\bibinfo
  {volume} {113}},\ \bibinfo {pages} {251302} (\bibinfo {year} {2014})},\
  \Eprint {http://arxiv.org/abs/1409.0549} {arXiv:1409.0549 [hep-ph]}
  \BibitemShut {NoStop}%
\bibitem [{\citenamefont {Poulin}\ \emph {et~al.}(2018)\citenamefont {Poulin},
  \citenamefont {Smith}, \citenamefont {Grin}, \citenamefont {Karwal},\ and\
  \citenamefont {Kamionkowski}}]{Poulin:2018dzj}%
  \BibitemOpen
  \bibfield  {author} {\bibinfo {author} {\bibfnamefont {V.}~\bibnamefont
  {Poulin}}, \bibinfo {author} {\bibfnamefont {T.~L.}\ \bibnamefont {Smith}},
  \bibinfo {author} {\bibfnamefont {D.}~\bibnamefont {Grin}}, \bibinfo {author}
  {\bibfnamefont {T.}~\bibnamefont {Karwal}}, \ and\ \bibinfo {author}
  {\bibfnamefont {M.}~\bibnamefont {Kamionkowski}},\ }\href {\doibase
  10.1103/PhysRevD.98.083525} {\bibfield  {journal} {\bibinfo  {journal} {Phys.
  Rev. D}\ }\textbf {\bibinfo {volume} {98}},\ \bibinfo {pages} {083525}
  (\bibinfo {year} {2018})},\ \Eprint {http://arxiv.org/abs/1806.10608}
  {arXiv:1806.10608 [astro-ph.CO]} \BibitemShut {NoStop}%
\bibitem [{\citenamefont {Poulin}\ \emph {et~al.}(2019)\citenamefont {Poulin},
  \citenamefont {Smith}, \citenamefont {Karwal},\ and\ \citenamefont
  {Kamionkowski}}]{Poulin:2018cxd}%
  \BibitemOpen
  \bibfield  {author} {\bibinfo {author} {\bibfnamefont {V.}~\bibnamefont
  {Poulin}}, \bibinfo {author} {\bibfnamefont {T.~L.}\ \bibnamefont {Smith}},
  \bibinfo {author} {\bibfnamefont {T.}~\bibnamefont {Karwal}}, \ and\ \bibinfo
  {author} {\bibfnamefont {M.}~\bibnamefont {Kamionkowski}},\ }\href {\doibase
  10.1103/PhysRevLett.122.221301} {\bibfield  {journal} {\bibinfo  {journal}
  {Phys. Rev. Lett.}\ }\textbf {\bibinfo {volume} {122}},\ \bibinfo {pages}
  {221301} (\bibinfo {year} {2019})},\ \Eprint
  {http://arxiv.org/abs/1811.04083} {arXiv:1811.04083 [astro-ph.CO]}
  \BibitemShut {NoStop}%
\bibitem [{\citenamefont {Peebles}\ and\ \citenamefont
  {Ratra}(1988)}]{Peebles:1987ek}%
  \BibitemOpen
  \bibfield  {author} {\bibinfo {author} {\bibfnamefont {P.~J.~E.}\
  \bibnamefont {Peebles}}\ and\ \bibinfo {author} {\bibfnamefont
  {B.}~\bibnamefont {Ratra}},\ }\href {\doibase 10.1086/185100} {\bibfield
  {journal} {\bibinfo  {journal} {Astrophys. J.}\ }\textbf {\bibinfo {volume}
  {325}},\ \bibinfo {pages} {L17} (\bibinfo {year} {1988})}\BibitemShut
  {NoStop}%
\bibitem [{\citenamefont {Ratra}\ and\ \citenamefont
  {Peebles}(1988)}]{Ratra:1987rm}%
  \BibitemOpen
  \bibfield  {author} {\bibinfo {author} {\bibfnamefont {B.}~\bibnamefont
  {Ratra}}\ and\ \bibinfo {author} {\bibfnamefont {P.~J.~E.}\ \bibnamefont
  {Peebles}},\ }\href {\doibase 10.1103/PhysRevD.37.3406} {\bibfield  {journal}
  {\bibinfo  {journal} {Phys. Rev.}\ }\textbf {\bibinfo {volume} {D37}},\
  \bibinfo {pages} {3406} (\bibinfo {year} {1988})}\BibitemShut {NoStop}%
\bibitem [{\citenamefont {Pettorino}\ \emph {et~al.}(2012)\citenamefont
  {Pettorino}, \citenamefont {Amendola}, \citenamefont {Baccigalupi},\ and\
  \citenamefont {Quercellini}}]{Pettorino:2012ts}%
  \BibitemOpen
  \bibfield  {author} {\bibinfo {author} {\bibfnamefont {V.}~\bibnamefont
  {Pettorino}}, \bibinfo {author} {\bibfnamefont {L.}~\bibnamefont {Amendola}},
  \bibinfo {author} {\bibfnamefont {C.}~\bibnamefont {Baccigalupi}}, \ and\
  \bibinfo {author} {\bibfnamefont {C.}~\bibnamefont {Quercellini}},\ }\href
  {\doibase 10.1103/PhysRevD.86.103507} {\bibfield  {journal} {\bibinfo
  {journal} {Phys. Rev. D}\ }\textbf {\bibinfo {volume} {86}},\ \bibinfo
  {pages} {103507} (\bibinfo {year} {2012})},\ \Eprint
  {http://arxiv.org/abs/1207.3293} {arXiv:1207.3293 [astro-ph.CO]} \BibitemShut
  {NoStop}%
\bibitem [{\citenamefont {Pettorino}(2013)}]{Pettorino:2013oxa}%
  \BibitemOpen
  \bibfield  {author} {\bibinfo {author} {\bibfnamefont {V.}~\bibnamefont
  {Pettorino}},\ }\href {\doibase 10.1103/PhysRevD.88.063519} {\bibfield
  {journal} {\bibinfo  {journal} {Phys. Rev. D}\ }\textbf {\bibinfo {volume}
  {88}},\ \bibinfo {pages} {063519} (\bibinfo {year} {2013})},\ \Eprint
  {http://arxiv.org/abs/1305.7457} {arXiv:1305.7457 [astro-ph.CO]} \BibitemShut
  {NoStop}%
\bibitem [{\citenamefont {Ade}\ \emph {et~al.}(2016)\citenamefont {Ade} \emph
  {et~al.}}]{Planck:2015bue}%
  \BibitemOpen
  \bibfield  {author} {\bibinfo {author} {\bibfnamefont {P.~A.~R.}\
  \bibnamefont {Ade}} \emph {et~al.} (\bibinfo {collaboration} {Planck}),\
  }\href {\doibase 10.1051/0004-6361/201525814} {\bibfield  {journal} {\bibinfo
   {journal} {Astron. Astrophys.}\ }\textbf {\bibinfo {volume} {594}},\
  \bibinfo {pages} {A14} (\bibinfo {year} {2016})},\ \Eprint
  {http://arxiv.org/abs/1502.01590} {arXiv:1502.01590 [astro-ph.CO]}
  \BibitemShut {NoStop}%
\bibitem [{\citenamefont {G\'omez-Valent}\ \emph {et~al.}(2020)\citenamefont
  {G\'omez-Valent}, \citenamefont {Pettorino},\ and\ \citenamefont
  {Amendola}}]{Gomez-Valent:2020mqn}%
  \BibitemOpen
  \bibfield  {author} {\bibinfo {author} {\bibfnamefont {A.}~\bibnamefont
  {G\'omez-Valent}}, \bibinfo {author} {\bibfnamefont {V.}~\bibnamefont
  {Pettorino}}, \ and\ \bibinfo {author} {\bibfnamefont {L.}~\bibnamefont
  {Amendola}},\ }\href {\doibase 10.1103/PhysRevD.101.123513} {\bibfield
  {journal} {\bibinfo  {journal} {Phys. Rev. D}\ }\textbf {\bibinfo {volume}
  {101}},\ \bibinfo {pages} {123513} (\bibinfo {year} {2020})},\ \Eprint
  {http://arxiv.org/abs/2004.00610} {arXiv:2004.00610 [astro-ph.CO]}
  \BibitemShut {NoStop}%
\bibitem [{\citenamefont {Wetterich}(1995)}]{Wetterich:1994bg}%
  \BibitemOpen
  \bibfield  {author} {\bibinfo {author} {\bibfnamefont {C.}~\bibnamefont
  {Wetterich}},\ }\href@noop {} {\bibfield  {journal} {\bibinfo  {journal}
  {Astron. Astrophys.}\ }\textbf {\bibinfo {volume} {301}},\ \bibinfo {pages}
  {321} (\bibinfo {year} {1995})},\ \Eprint
  {http://arxiv.org/abs/hep-th/9408025} {arXiv:hep-th/9408025} \BibitemShut
  {NoStop}%
\bibitem [{\citenamefont {Amendola}(2000)}]{Amendola:1999er}%
  \BibitemOpen
  \bibfield  {author} {\bibinfo {author} {\bibfnamefont {L.}~\bibnamefont
  {Amendola}},\ }\href {\doibase 10.1103/PhysRevD.62.043511} {\bibfield
  {journal} {\bibinfo  {journal} {Phys. Rev. D}\ }\textbf {\bibinfo {volume}
  {62}},\ \bibinfo {pages} {043511} (\bibinfo {year} {2000})},\ \Eprint
  {http://arxiv.org/abs/astro-ph/9908023} {arXiv:astro-ph/9908023} \BibitemShut
  {NoStop}%
\bibitem [{\citenamefont {Brans}\ and\ \citenamefont
  {Dicke}(1961)}]{BransDicke1961}%
  \BibitemOpen
  \bibfield  {author} {\bibinfo {author} {\bibfnamefont {C.}~\bibnamefont
  {Brans}}\ and\ \bibinfo {author} {\bibfnamefont {R.}~\bibnamefont {Dicke}},\
  }\href@noop {} {\bibfield  {journal} {\bibinfo  {journal} {Phys. Rev}\
  }\textbf {\bibinfo {volume} {124}},\ \bibinfo {pages} {925} (\bibinfo {year}
  {1961})}\BibitemShut {NoStop}%
\bibitem [{\citenamefont {Dicke}(1962)}]{dicke1962physical}%
  \BibitemOpen
  \bibfield  {author} {\bibinfo {author} {\bibfnamefont {R.}~\bibnamefont
  {Dicke}},\ }\href@noop {} {\bibfield  {journal} {\bibinfo  {journal} {Phys.
  Rev.}\ }\textbf {\bibinfo {volume} {125}},\ \bibinfo {pages} {2163} (\bibinfo
  {year} {1962})}\BibitemShut {NoStop}%
\bibitem [{\citenamefont {Brans}(1962)}]{brans1962mach}%
  \BibitemOpen
  \bibfield  {author} {\bibinfo {author} {\bibfnamefont {C.}~\bibnamefont
  {Brans}},\ }\href@noop {} {\bibfield  {journal} {\bibinfo  {journal} {Phys.
  Rev.}\ }\textbf {\bibinfo {volume} {125}},\ \bibinfo {pages} {2194} (\bibinfo
  {year} {1962})}\BibitemShut {NoStop}%
\bibitem [{\citenamefont {Avilez}\ and\ \citenamefont
  {Skordis}(2014)}]{Avilez:2013dxa}%
  \BibitemOpen
  \bibfield  {author} {\bibinfo {author} {\bibfnamefont {A.}~\bibnamefont
  {Avilez}}\ and\ \bibinfo {author} {\bibfnamefont {C.}~\bibnamefont
  {Skordis}},\ }\href {\doibase 10.1103/PhysRevLett.113.011101} {\bibfield
  {journal} {\bibinfo  {journal} {Phys. Rev. Lett.}\ }\textbf {\bibinfo
  {volume} {113}},\ \bibinfo {pages} {011101} (\bibinfo {year} {2014})},\
  \bibinfo {note} {arXiv:1303.4330},\ \Eprint {http://arxiv.org/abs/1303.4330}
  {arXiv:1303.4330 [astro-ph.CO]} \BibitemShut {NoStop}%
\bibitem [{\citenamefont {de~Cruz~P\'erez}\ and\ \citenamefont
  {Sol\`a~Peracaula}(2018)}]{deCruzPerez:2018cjx}%
  \BibitemOpen
  \bibfield  {author} {\bibinfo {author} {\bibfnamefont {J.}~\bibnamefont
  {de~Cruz~P\'erez}}\ and\ \bibinfo {author} {\bibfnamefont {J.}~\bibnamefont
  {Sol\`a~Peracaula}},\ }\href {\doibase 10.1142/S0217732318502280} {\bibfield
  {journal} {\bibinfo  {journal} {Mod. Phys. Lett. A}\ }\textbf {\bibinfo
  {volume} {33}},\ \bibinfo {pages} {1850228} (\bibinfo {year} {2018})},\
  \Eprint {http://arxiv.org/abs/1809.03329} {arXiv:1809.03329 [gr-qc]}
  \BibitemShut {NoStop}%
\bibitem [{\citenamefont {Sol\`a~Peracaula}\ \emph {et~al.}(2019)\citenamefont
  {Sol\`a~Peracaula}, \citenamefont {G\'omez-Valent}, \citenamefont
  {de~Cruz~P\'erez},\ and\ \citenamefont
  {Moreno-Pulido}}]{SolaPeracaula:2019zsl}%
  \BibitemOpen
  \bibfield  {author} {\bibinfo {author} {\bibfnamefont {J.}~\bibnamefont
  {Sol\`a~Peracaula}}, \bibinfo {author} {\bibfnamefont {A.}~\bibnamefont
  {G\'omez-Valent}}, \bibinfo {author} {\bibfnamefont {J.}~\bibnamefont
  {de~Cruz~P\'erez}}, \ and\ \bibinfo {author} {\bibfnamefont {C.}~\bibnamefont
  {Moreno-Pulido}},\ }\href {\doibase 10.3847/2041-8213/ab53e9} {\bibfield
  {journal} {\bibinfo  {journal} {Astrophys. J. Lett.}\ }\textbf {\bibinfo
  {volume} {886}},\ \bibinfo {pages} {L6} (\bibinfo {year} {2019})},\ \Eprint
  {http://arxiv.org/abs/1909.02554} {arXiv:1909.02554 [astro-ph.CO]}
  \BibitemShut {NoStop}%
\bibitem [{\citenamefont {Sol\`a~Peracaula}\ \emph {et~al.}(2020)\citenamefont
  {Sol\`a~Peracaula}, \citenamefont {G\'omez-Valent}, \citenamefont
  {de~Cruz~P\'erez},\ and\ \citenamefont
  {Moreno-Pulido}}]{SolaPeracaula:2020vpg}%
  \BibitemOpen
  \bibfield  {author} {\bibinfo {author} {\bibfnamefont {J.}~\bibnamefont
  {Sol\`a~Peracaula}}, \bibinfo {author} {\bibfnamefont {A.}~\bibnamefont
  {G\'omez-Valent}}, \bibinfo {author} {\bibfnamefont {J.}~\bibnamefont
  {de~Cruz~P\'erez}}, \ and\ \bibinfo {author} {\bibfnamefont {C.}~\bibnamefont
  {Moreno-Pulido}},\ }\href {\doibase 10.1088/1361-6382/abbc43} {\bibfield
  {journal} {\bibinfo  {journal} {Class. Quant. Grav.}\ }\textbf {\bibinfo
  {volume} {37}},\ \bibinfo {pages} {245003} (\bibinfo {year} {2020})},\
  \Eprint {http://arxiv.org/abs/2006.04273} {arXiv:2006.04273 [astro-ph.CO]}
  \BibitemShut {NoStop}%
\bibitem [{\citenamefont {Joudaki}\ \emph {et~al.}(2022)\citenamefont
  {Joudaki}, \citenamefont {Ferreira}, \citenamefont {Lima},\ and\
  \citenamefont {Winther}}]{Joudaki:2020shz}%
  \BibitemOpen
  \bibfield  {author} {\bibinfo {author} {\bibfnamefont {S.}~\bibnamefont
  {Joudaki}}, \bibinfo {author} {\bibfnamefont {P.~G.}\ \bibnamefont
  {Ferreira}}, \bibinfo {author} {\bibfnamefont {N.~A.}\ \bibnamefont {Lima}},
  \ and\ \bibinfo {author} {\bibfnamefont {H.~A.}\ \bibnamefont {Winther}},\
  }\href {\doibase 10.1103/PhysRevD.105.043522} {\bibfield  {journal} {\bibinfo
   {journal} {Phys. Rev. D}\ }\textbf {\bibinfo {volume} {105}},\ \bibinfo
  {pages} {043522} (\bibinfo {year} {2022})},\ \Eprint
  {http://arxiv.org/abs/2010.15278} {arXiv:2010.15278 [astro-ph.CO]}
  \BibitemShut {NoStop}%
\bibitem [{\citenamefont {Sol\`a~Peracaula}\ \emph {et~al.}(2021)\citenamefont
  {Sol\`a~Peracaula}, \citenamefont {G\'omez-Valent}, \citenamefont
  {de~Cruz~P\'erez},\ and\ \citenamefont
  {Moreno-Pulido}}]{SolaPeracaula:2021gxi}%
  \BibitemOpen
  \bibfield  {author} {\bibinfo {author} {\bibfnamefont {J.}~\bibnamefont
  {Sol\`a~Peracaula}}, \bibinfo {author} {\bibfnamefont {A.}~\bibnamefont
  {G\'omez-Valent}}, \bibinfo {author} {\bibfnamefont {J.}~\bibnamefont
  {de~Cruz~P\'erez}}, \ and\ \bibinfo {author} {\bibfnamefont {C.}~\bibnamefont
  {Moreno-Pulido}},\ }\href {\doibase 10.1209/0295-5075/134/19001} {\bibfield
  {journal} {\bibinfo  {journal} {EPL}\ }\textbf {\bibinfo {volume} {134}},\
  \bibinfo {pages} {19001} (\bibinfo {year} {2021})},\ \Eprint
  {http://arxiv.org/abs/2102.12758} {arXiv:2102.12758 [astro-ph.CO]}
  \BibitemShut {NoStop}%
\bibitem [{\citenamefont {Ade}\ \emph {et~al.}(2014)\citenamefont {Ade} \emph
  {et~al.}}]{Planck:2013nga}%
  \BibitemOpen
  \bibfield  {author} {\bibinfo {author} {\bibfnamefont {P.~A.~R.}\
  \bibnamefont {Ade}} \emph {et~al.} (\bibinfo {collaboration} {Planck}),\
  }\href {\doibase 10.1051/0004-6361/201323003} {\bibfield  {journal} {\bibinfo
   {journal} {Astron. Astrophys.}\ }\textbf {\bibinfo {volume} {566}},\
  \bibinfo {pages} {A54} (\bibinfo {year} {2014})},\ \Eprint
  {http://arxiv.org/abs/1311.1657} {arXiv:1311.1657 [astro-ph.CO]} \BibitemShut
  {NoStop}%
\bibitem [{\citenamefont {Aghanim}\ \emph {et~al.}(2020)\citenamefont {Aghanim}
  \emph {et~al.}}]{Planck:2018vyg}%
  \BibitemOpen
  \bibfield  {author} {\bibinfo {author} {\bibfnamefont {N.}~\bibnamefont
  {Aghanim}} \emph {et~al.} (\bibinfo {collaboration} {Planck}),\ }\href
  {\doibase 10.1051/0004-6361/201833910} {\bibfield  {journal} {\bibinfo
  {journal} {Astron. Astrophys.}\ }\textbf {\bibinfo {volume} {641}},\ \bibinfo
  {pages} {A6} (\bibinfo {year} {2020})},\ \Eprint
  {http://arxiv.org/abs/1807.06209} {arXiv:1807.06209 [astro-ph.CO]}
  \BibitemShut {NoStop}%
\bibitem [{\citenamefont {Herold}\ \emph {et~al.}(2022)\citenamefont {Herold},
  \citenamefont {Ferreira},\ and\ \citenamefont {Komatsu}}]{Herold:2021ksg}%
  \BibitemOpen
  \bibfield  {author} {\bibinfo {author} {\bibfnamefont {L.}~\bibnamefont
  {Herold}}, \bibinfo {author} {\bibfnamefont {E.~G.~M.}\ \bibnamefont
  {Ferreira}}, \ and\ \bibinfo {author} {\bibfnamefont {E.}~\bibnamefont
  {Komatsu}},\ }\href {\doibase 10.3847/2041-8213/ac63a3} {\bibfield  {journal}
  {\bibinfo  {journal} {Astrophys. J. Lett.}\ }\textbf {\bibinfo {volume}
  {929}},\ \bibinfo {pages} {L16} (\bibinfo {year} {2022})},\ \Eprint
  {http://arxiv.org/abs/2112.12140} {arXiv:2112.12140 [astro-ph.CO]}
  \BibitemShut {NoStop}%
\bibitem [{\citenamefont {Smith}\ \emph {et~al.}(2021)\citenamefont {Smith},
  \citenamefont {Poulin}, \citenamefont {Bernal}, \citenamefont {Boddy},
  \citenamefont {Kamionkowski},\ and\ \citenamefont {Murgia}}]{Smith:2020rxx}%
  \BibitemOpen
  \bibfield  {author} {\bibinfo {author} {\bibfnamefont {T.~L.}\ \bibnamefont
  {Smith}}, \bibinfo {author} {\bibfnamefont {V.}~\bibnamefont {Poulin}},
  \bibinfo {author} {\bibfnamefont {J.~L.}\ \bibnamefont {Bernal}}, \bibinfo
  {author} {\bibfnamefont {K.~K.}\ \bibnamefont {Boddy}}, \bibinfo {author}
  {\bibfnamefont {M.}~\bibnamefont {Kamionkowski}}, \ and\ \bibinfo {author}
  {\bibfnamefont {R.}~\bibnamefont {Murgia}},\ }\href {\doibase
  10.1103/PhysRevD.103.123542} {\bibfield  {journal} {\bibinfo  {journal}
  {Phys. Rev. D}\ }\textbf {\bibinfo {volume} {103}},\ \bibinfo {pages}
  {123542} (\bibinfo {year} {2021})},\ \Eprint
  {http://arxiv.org/abs/2009.10740} {arXiv:2009.10740 [astro-ph.CO]}
  \BibitemShut {NoStop}%
\bibitem [{\citenamefont {Ivanov}\ \emph {et~al.}(2020)\citenamefont {Ivanov},
  \citenamefont {McDonough}, \citenamefont {Hill}, \citenamefont {Simonovi\'c},
  \citenamefont {Toomey}, \citenamefont {Alexander},\ and\ \citenamefont
  {Zaldarriaga}}]{Ivanov:2020ril}%
  \BibitemOpen
  \bibfield  {author} {\bibinfo {author} {\bibfnamefont {M.~M.}\ \bibnamefont
  {Ivanov}}, \bibinfo {author} {\bibfnamefont {E.}~\bibnamefont {McDonough}},
  \bibinfo {author} {\bibfnamefont {J.~C.}\ \bibnamefont {Hill}}, \bibinfo
  {author} {\bibfnamefont {M.}~\bibnamefont {Simonovi\'c}}, \bibinfo {author}
  {\bibfnamefont {M.~W.}\ \bibnamefont {Toomey}}, \bibinfo {author}
  {\bibfnamefont {S.}~\bibnamefont {Alexander}}, \ and\ \bibinfo {author}
  {\bibfnamefont {M.}~\bibnamefont {Zaldarriaga}},\ }\href {\doibase
  10.1103/PhysRevD.102.103502} {\bibfield  {journal} {\bibinfo  {journal}
  {Phys. Rev. D}\ }\textbf {\bibinfo {volume} {102}},\ \bibinfo {pages}
  {103502} (\bibinfo {year} {2020})},\ \Eprint
  {http://arxiv.org/abs/2006.11235} {arXiv:2006.11235 [astro-ph.CO]}
  \BibitemShut {NoStop}%
\bibitem [{\citenamefont {Akaike}(1974)}]{Akaike}%
  \BibitemOpen
  \bibfield  {author} {\bibinfo {author} {\bibfnamefont {H.}~\bibnamefont
  {Akaike}},\ }\href {\doibase 10.1109/TAC.1974.1100705} {\bibfield  {journal}
  {\bibinfo  {journal} {IEEE Trans. Autom. Control}\ }\textbf {\bibinfo
  {volume} {19}},\ \bibinfo {pages} {716} (\bibinfo {year} {1974})}\BibitemShut
  {NoStop}%
\bibitem [{\citenamefont {Schwarz}(1978)}]{Schwarz1978}%
  \BibitemOpen
  \bibfield  {author} {\bibinfo {author} {\bibfnamefont {G.}~\bibnamefont
  {Schwarz}},\ }\href {\doibase 10.1214/aos/1176344136} {\bibfield  {journal}
  {\bibinfo  {journal} {Ann. Stat.}\ }\textbf {\bibinfo {volume} {6}},\
  \bibinfo {pages} {461} (\bibinfo {year} {1978})}\BibitemShut {NoStop}%
\bibitem [{\citenamefont {Kass}\ and\ \citenamefont
  {Raftery}(1995)}]{KassRaftery1995}%
  \BibitemOpen
  \bibfield  {author} {\bibinfo {author} {\bibfnamefont {R.~E.}\ \bibnamefont
  {Kass}}\ and\ \bibinfo {author} {\bibfnamefont {A.~E.}\ \bibnamefont
  {Raftery}},\ }\href {\doibase 10.2307/2291091} {\bibfield  {journal}
  {\bibinfo  {journal} {J. Amer. Statist. Assoc.}\ }\textbf {\bibinfo {volume}
  {90}},\ \bibinfo {pages} {773} (\bibinfo {year} {1995})}\BibitemShut
  {NoStop}%
\bibitem [{\citenamefont {Feldman}\ and\ \citenamefont
  {Cousins}(1998)}]{Feldman:1997qc}%
  \BibitemOpen
  \bibfield  {author} {\bibinfo {author} {\bibfnamefont {G.~J.}\ \bibnamefont
  {Feldman}}\ and\ \bibinfo {author} {\bibfnamefont {R.~D.}\ \bibnamefont
  {Cousins}},\ }\href {\doibase 10.1103/PhysRevD.57.3873} {\bibfield  {journal}
  {\bibinfo  {journal} {Phys. Rev. D}\ }\textbf {\bibinfo {volume} {57}},\
  \bibinfo {pages} {3873} (\bibinfo {year} {1998})},\ \Eprint
  {http://arxiv.org/abs/physics/9711021} {arXiv:physics/9711021} \BibitemShut
  {NoStop}%
\bibitem [{\citenamefont {Scolnic}\ \emph {et~al.}(2018)\citenamefont {Scolnic}
  \emph {et~al.}}]{Scolnic:2017caz}%
  \BibitemOpen
  \bibfield  {author} {\bibinfo {author} {\bibfnamefont {D.~M.}\ \bibnamefont
  {Scolnic}} \emph {et~al.},\ }\href {\doibase 10.3847/1538-4357/aab9bb}
  {\bibfield  {journal} {\bibinfo  {journal} {Astrophys. J.}\ }\textbf
  {\bibinfo {volume} {859}},\ \bibinfo {pages} {101} (\bibinfo {year}
  {2018})},\ \Eprint {http://arxiv.org/abs/1710.00845} {arXiv:1710.00845
  [astro-ph.CO]} \BibitemShut {NoStop}%
\bibitem [{\citenamefont {Carter}\ \emph {et~al.}(2018)\citenamefont {Carter},
  \citenamefont {Beutler}, \citenamefont {Percival}, \citenamefont {Blake},
  \citenamefont {Koda},\ and\ \citenamefont {Ross}}]{Carter:2018vce}%
  \BibitemOpen
  \bibfield  {author} {\bibinfo {author} {\bibfnamefont {P.}~\bibnamefont
  {Carter}}, \bibinfo {author} {\bibfnamefont {F.}~\bibnamefont {Beutler}},
  \bibinfo {author} {\bibfnamefont {W.~J.}\ \bibnamefont {Percival}}, \bibinfo
  {author} {\bibfnamefont {C.}~\bibnamefont {Blake}}, \bibinfo {author}
  {\bibfnamefont {J.}~\bibnamefont {Koda}}, \ and\ \bibinfo {author}
  {\bibfnamefont {A.~J.}\ \bibnamefont {Ross}},\ }\href {\doibase
  10.1093/mnras/sty2405} {\bibfield  {journal} {\bibinfo  {journal} {Mon. Not.
  Roy. Astron. Soc.}\ }\textbf {\bibinfo {volume} {481}},\ \bibinfo {pages}
  {2371} (\bibinfo {year} {2018})},\ \Eprint {http://arxiv.org/abs/1803.01746}
  {arXiv:1803.01746 [astro-ph.CO]} \BibitemShut {NoStop}%
\bibitem [{\citenamefont {Beutler}\ \emph {et~al.}(2011)\citenamefont
  {Beutler}, \citenamefont {Blake}, \citenamefont {Colless}, \citenamefont
  {Jones}, \citenamefont {Staveley-Smith}, \citenamefont {Campbell},
  \citenamefont {Parker}, \citenamefont {Saunders},\ and\ \citenamefont
  {Watson}}]{Beutler:2011hx}%
  \BibitemOpen
  \bibfield  {author} {\bibinfo {author} {\bibfnamefont {F.}~\bibnamefont
  {Beutler}}, \bibinfo {author} {\bibfnamefont {C.}~\bibnamefont {Blake}},
  \bibinfo {author} {\bibfnamefont {M.}~\bibnamefont {Colless}}, \bibinfo
  {author} {\bibfnamefont {D.~H.}\ \bibnamefont {Jones}}, \bibinfo {author}
  {\bibfnamefont {L.}~\bibnamefont {Staveley-Smith}}, \bibinfo {author}
  {\bibfnamefont {L.}~\bibnamefont {Campbell}}, \bibinfo {author}
  {\bibfnamefont {Q.}~\bibnamefont {Parker}}, \bibinfo {author} {\bibfnamefont
  {W.}~\bibnamefont {Saunders}}, \ and\ \bibinfo {author} {\bibfnamefont
  {F.}~\bibnamefont {Watson}},\ }\href {\doibase
  10.1111/j.1365-2966.2011.19250.x} {\bibfield  {journal} {\bibinfo  {journal}
  {Mon. Not. Roy. Astron. Soc.}\ }\textbf {\bibinfo {volume} {416}},\ \bibinfo
  {pages} {3017} (\bibinfo {year} {2011})},\ \Eprint
  {http://arxiv.org/abs/1106.3366} {arXiv:1106.3366 [astro-ph.CO]} \BibitemShut
  {NoStop}%
\bibitem [{\citenamefont {Ross}\ \emph {et~al.}(2015)\citenamefont {Ross},
  \citenamefont {Samushia}, \citenamefont {Howlett}, \citenamefont {Percival},
  \citenamefont {Burden},\ and\ \citenamefont {Manera}}]{Ross:2014qpa}%
  \BibitemOpen
  \bibfield  {author} {\bibinfo {author} {\bibfnamefont {A.~J.}\ \bibnamefont
  {Ross}}, \bibinfo {author} {\bibfnamefont {L.}~\bibnamefont {Samushia}},
  \bibinfo {author} {\bibfnamefont {C.}~\bibnamefont {Howlett}}, \bibinfo
  {author} {\bibfnamefont {W.~J.}\ \bibnamefont {Percival}}, \bibinfo {author}
  {\bibfnamefont {A.}~\bibnamefont {Burden}}, \ and\ \bibinfo {author}
  {\bibfnamefont {M.}~\bibnamefont {Manera}},\ }\href {\doibase
  10.1093/mnras/stv154} {\bibfield  {journal} {\bibinfo  {journal} {Mon. Not.
  Roy. Astron. Soc.}\ }\textbf {\bibinfo {volume} {449}},\ \bibinfo {pages}
  {835} (\bibinfo {year} {2015})},\ \Eprint {http://arxiv.org/abs/1409.3242}
  {arXiv:1409.3242 [astro-ph.CO]} \BibitemShut {NoStop}%
\bibitem [{\citenamefont {Gil-Mar\'\i{}n}\ \emph {et~al.}(2017)\citenamefont
  {Gil-Mar\'\i{}n}, \citenamefont {Percival}, \citenamefont {Verde},
  \citenamefont {Brownstein}, \citenamefont {Chuang}, \citenamefont {Kitaura},
  \citenamefont {Rodr\'\i{}guez-Torres},\ and\ \citenamefont
  {Olmstead}}]{Gil-Marin:2016wya}%
  \BibitemOpen
  \bibfield  {author} {\bibinfo {author} {\bibfnamefont {H.}~\bibnamefont
  {Gil-Mar\'\i{}n}}, \bibinfo {author} {\bibfnamefont {W.~J.}\ \bibnamefont
  {Percival}}, \bibinfo {author} {\bibfnamefont {L.}~\bibnamefont {Verde}},
  \bibinfo {author} {\bibfnamefont {J.~R.}\ \bibnamefont {Brownstein}},
  \bibinfo {author} {\bibfnamefont {C.-H.}\ \bibnamefont {Chuang}}, \bibinfo
  {author} {\bibfnamefont {F.-S.}\ \bibnamefont {Kitaura}}, \bibinfo {author}
  {\bibfnamefont {S.~A.}\ \bibnamefont {Rodr\'\i{}guez-Torres}}, \ and\
  \bibinfo {author} {\bibfnamefont {M.~D.}\ \bibnamefont {Olmstead}},\ }\href
  {\doibase 10.1093/mnras/stw2679} {\bibfield  {journal} {\bibinfo  {journal}
  {Mon. Not. Roy. Astron. Soc.}\ }\textbf {\bibinfo {volume} {465}},\ \bibinfo
  {pages} {1757} (\bibinfo {year} {2017})},\ \Eprint
  {http://arxiv.org/abs/1606.00439} {arXiv:1606.00439 [astro-ph.CO]}
  \BibitemShut {NoStop}%
\bibitem [{\citenamefont {Kazin}\ \emph {et~al.}(2014)\citenamefont {Kazin}
  \emph {et~al.}}]{Kazin:2014qga}%
  \BibitemOpen
  \bibfield  {author} {\bibinfo {author} {\bibfnamefont {E.~A.}\ \bibnamefont
  {Kazin}} \emph {et~al.},\ }\href {\doibase 10.1093/mnras/stu778} {\bibfield
  {journal} {\bibinfo  {journal} {Mon. Not. Roy. Astron. Soc.}\ }\textbf
  {\bibinfo {volume} {441}},\ \bibinfo {pages} {3524} (\bibinfo {year}
  {2014})},\ \Eprint {http://arxiv.org/abs/1401.0358} {arXiv:1401.0358
  [astro-ph.CO]} \BibitemShut {NoStop}%
\bibitem [{\citenamefont {Abbott}\ \emph {et~al.}(2019)\citenamefont {Abbott}
  \emph {et~al.}}]{DES:2017rfo}%
  \BibitemOpen
  \bibfield  {author} {\bibinfo {author} {\bibfnamefont {T.~M.~C.}\
  \bibnamefont {Abbott}} \emph {et~al.} (\bibinfo {collaboration} {DES}),\
  }\href {\doibase 10.1093/mnras/sty3351} {\bibfield  {journal} {\bibinfo
  {journal} {Mon. Not. Roy. Astron. Soc.}\ }\textbf {\bibinfo {volume} {483}},\
  \bibinfo {pages} {4866} (\bibinfo {year} {2019})},\ \Eprint
  {http://arxiv.org/abs/1712.06209} {arXiv:1712.06209 [astro-ph.CO]}
  \BibitemShut {NoStop}%
\bibitem [{\citenamefont {Neveux}\ \emph {et~al.}(2020)\citenamefont {Neveux}
  \emph {et~al.}}]{Neveux:2020voa}%
  \BibitemOpen
  \bibfield  {author} {\bibinfo {author} {\bibfnamefont {R.}~\bibnamefont
  {Neveux}} \emph {et~al.},\ }\href {\doibase 10.1093/mnras/staa2780}
  {\bibfield  {journal} {\bibinfo  {journal} {Mon. Not. Roy. Astron. Soc.}\
  }\textbf {\bibinfo {volume} {499}},\ \bibinfo {pages} {210} (\bibinfo {year}
  {2020})},\ \Eprint {http://arxiv.org/abs/2007.08999} {arXiv:2007.08999
  [astro-ph.CO]} \BibitemShut {NoStop}%
\bibitem [{\citenamefont {du~Mas~des Bourboux}\ \emph
  {et~al.}(2020)\citenamefont {du~Mas~des Bourboux} \emph
  {et~al.}}]{duMasdesBourboux:2020pck}%
  \BibitemOpen
  \bibfield  {author} {\bibinfo {author} {\bibfnamefont {H.}~\bibnamefont
  {du~Mas~des Bourboux}} \emph {et~al.},\ }\href {\doibase
  10.3847/1538-4357/abb085} {\bibfield  {journal} {\bibinfo  {journal}
  {Astrophys. J.}\ }\textbf {\bibinfo {volume} {901}},\ \bibinfo {pages} {153}
  (\bibinfo {year} {2020})},\ \Eprint {http://arxiv.org/abs/2007.08995}
  {arXiv:2007.08995 [astro-ph.CO]} \BibitemShut {NoStop}%
\bibitem [{\citenamefont {Aubourg}\ \emph {et~al.}(2015)\citenamefont {Aubourg}
  \emph {et~al.}}]{Aubourg:2014yra}%
  \BibitemOpen
  \bibfield  {author} {\bibinfo {author} {\bibfnamefont {E.}~\bibnamefont
  {Aubourg}} \emph {et~al.},\ }\href {\doibase 10.1103/PhysRevD.92.123516}
  {\bibfield  {journal} {\bibinfo  {journal} {Phys. Rev. D}\ }\textbf {\bibinfo
  {volume} {92}},\ \bibinfo {pages} {123516} (\bibinfo {year} {2015})},\
  \Eprint {http://arxiv.org/abs/1411.1074} {arXiv:1411.1074 [astro-ph.CO]}
  \BibitemShut {NoStop}%
\bibitem [{\citenamefont {Cuesta}\ \emph {et~al.}(2015)\citenamefont {Cuesta},
  \citenamefont {Verde}, \citenamefont {Riess},\ and\ \citenamefont
  {Jimenez}}]{Cuesta:2014asa}%
  \BibitemOpen
  \bibfield  {author} {\bibinfo {author} {\bibfnamefont {A.~J.}\ \bibnamefont
  {Cuesta}}, \bibinfo {author} {\bibfnamefont {L.}~\bibnamefont {Verde}},
  \bibinfo {author} {\bibfnamefont {A.}~\bibnamefont {Riess}}, \ and\ \bibinfo
  {author} {\bibfnamefont {R.}~\bibnamefont {Jimenez}},\ }\href {\doibase
  10.1093/mnras/stv261} {\bibfield  {journal} {\bibinfo  {journal} {Mon. Not.
  Roy. Astron. Soc.}\ }\textbf {\bibinfo {volume} {448}},\ \bibinfo {pages}
  {3463} (\bibinfo {year} {2015})},\ \Eprint {http://arxiv.org/abs/1411.1094}
  {arXiv:1411.1094 [astro-ph.CO]} \BibitemShut {NoStop}%
\bibitem [{\citenamefont {Feeney}\ \emph {et~al.}(2019)\citenamefont {Feeney},
  \citenamefont {Peiris}, \citenamefont {Williamson}, \citenamefont {Nissanke},
  \citenamefont {Mortlock}, \citenamefont {Alsing},\ and\ \citenamefont
  {Scolnic}}]{Feeney:2018mkj}%
  \BibitemOpen
  \bibfield  {author} {\bibinfo {author} {\bibfnamefont {S.~M.}\ \bibnamefont
  {Feeney}}, \bibinfo {author} {\bibfnamefont {H.~V.}\ \bibnamefont {Peiris}},
  \bibinfo {author} {\bibfnamefont {A.~R.}\ \bibnamefont {Williamson}},
  \bibinfo {author} {\bibfnamefont {S.~M.}\ \bibnamefont {Nissanke}}, \bibinfo
  {author} {\bibfnamefont {D.~J.}\ \bibnamefont {Mortlock}}, \bibinfo {author}
  {\bibfnamefont {J.}~\bibnamefont {Alsing}}, \ and\ \bibinfo {author}
  {\bibfnamefont {D.}~\bibnamefont {Scolnic}},\ }\href {\doibase
  10.1103/PhysRevLett.122.061105} {\bibfield  {journal} {\bibinfo  {journal}
  {Phys. Rev. Lett.}\ }\textbf {\bibinfo {volume} {122}},\ \bibinfo {pages}
  {061105} (\bibinfo {year} {2019})},\ \Eprint
  {http://arxiv.org/abs/1802.03404} {arXiv:1802.03404 [astro-ph.CO]}
  \BibitemShut {NoStop}%
\bibitem [{\citenamefont {Camarena}\ and\ \citenamefont
  {Marra}(2020{\natexlab{a}})}]{Camarena:2019rmj}%
  \BibitemOpen
  \bibfield  {author} {\bibinfo {author} {\bibfnamefont {D.}~\bibnamefont
  {Camarena}}\ and\ \bibinfo {author} {\bibfnamefont {V.}~\bibnamefont
  {Marra}},\ }\href {\doibase 10.1093/mnras/staa770} {\bibfield  {journal}
  {\bibinfo  {journal} {Mon. Not. Roy. Astron. Soc.}\ }\textbf {\bibinfo
  {volume} {495}},\ \bibinfo {pages} {2630} (\bibinfo {year}
  {2020}{\natexlab{a}})},\ \Eprint {http://arxiv.org/abs/1910.14125}
  {arXiv:1910.14125 [astro-ph.CO]} \BibitemShut {NoStop}%
\bibitem [{\citenamefont {Riess}\ \emph {et~al.}(2022)\citenamefont {Riess}
  \emph {et~al.}}]{Riess:2021jrx}%
  \BibitemOpen
  \bibfield  {author} {\bibinfo {author} {\bibfnamefont {A.~G.}\ \bibnamefont
  {Riess}} \emph {et~al.},\ }\href {\doibase 10.3847/2041-8213/ac5c5b}
  {\bibfield  {journal} {\bibinfo  {journal} {Astrophys. J. Lett.}\ }\textbf
  {\bibinfo {volume} {934}},\ \bibinfo {pages} {L7} (\bibinfo {year} {2022})},\
  \Eprint {http://arxiv.org/abs/2112.04510} {arXiv:2112.04510 [astro-ph.CO]}
  \BibitemShut {NoStop}%
\bibitem [{\citenamefont {Camarena}\ and\ \citenamefont
  {Marra}(2020{\natexlab{b}})}]{Camarena:2019moy}%
  \BibitemOpen
  \bibfield  {author} {\bibinfo {author} {\bibfnamefont {D.}~\bibnamefont
  {Camarena}}\ and\ \bibinfo {author} {\bibfnamefont {V.}~\bibnamefont
  {Marra}},\ }\href {\doibase 10.1103/PhysRevResearch.2.013028} {\bibfield
  {journal} {\bibinfo  {journal} {Phys. Rev. Res.}\ }\textbf {\bibinfo {volume}
  {2}},\ \bibinfo {pages} {013028} (\bibinfo {year} {2020}{\natexlab{b}})},\
  \Eprint {http://arxiv.org/abs/1906.11814} {arXiv:1906.11814 [astro-ph.CO]}
  \BibitemShut {NoStop}%
\bibitem [{\citenamefont {Efstathiou}(2021)}]{Efstathiou:2021ocp}%
  \BibitemOpen
  \bibfield  {author} {\bibinfo {author} {\bibfnamefont {G.}~\bibnamefont
  {Efstathiou}},\ }\href {\doibase 10.1093/mnras/stab1588} {\bibfield
  {journal} {\bibinfo  {journal} {Mon. Not. Roy. Astron. Soc.}\ }\textbf
  {\bibinfo {volume} {505}},\ \bibinfo {pages} {3866} (\bibinfo {year}
  {2021})},\ \Eprint {http://arxiv.org/abs/2103.08723} {arXiv:2103.08723
  [astro-ph.CO]} \BibitemShut {NoStop}%
\bibitem [{\citenamefont {Blas}\ \emph {et~al.}(2011)\citenamefont {Blas},
  \citenamefont {Lesgourgues},\ and\ \citenamefont {Tram}}]{Blas:2011rf}%
  \BibitemOpen
  \bibfield  {author} {\bibinfo {author} {\bibfnamefont {D.}~\bibnamefont
  {Blas}}, \bibinfo {author} {\bibfnamefont {J.}~\bibnamefont {Lesgourgues}}, \
  and\ \bibinfo {author} {\bibfnamefont {T.}~\bibnamefont {Tram}},\ }\href
  {\doibase 10.1088/1475-7516/2011/07/034} {\bibfield  {journal} {\bibinfo
  {journal} {JCAP}\ }\textbf {\bibinfo {volume} {07}},\ \bibinfo {pages} {034}
  (\bibinfo {year} {2011})},\ \Eprint {http://arxiv.org/abs/1104.2933}
  {arXiv:1104.2933 [astro-ph.CO]} \BibitemShut {NoStop}%
\bibitem [{\citenamefont {Audren}\ \emph {et~al.}(2013)\citenamefont {Audren},
  \citenamefont {Lesgourgues}, \citenamefont {Benabed},\ and\ \citenamefont
  {Prunet}}]{Audren:2012wb}%
  \BibitemOpen
  \bibfield  {author} {\bibinfo {author} {\bibfnamefont {B.}~\bibnamefont
  {Audren}}, \bibinfo {author} {\bibfnamefont {J.}~\bibnamefont {Lesgourgues}},
  \bibinfo {author} {\bibfnamefont {K.}~\bibnamefont {Benabed}}, \ and\
  \bibinfo {author} {\bibfnamefont {S.}~\bibnamefont {Prunet}},\ }\href
  {\doibase 10.1088/1475-7516/2013/02/001} {\bibfield  {journal} {\bibinfo
  {journal} {JCAP}\ }\textbf {\bibinfo {volume} {02}},\ \bibinfo {pages} {001}
  (\bibinfo {year} {2013})},\ \Eprint {http://arxiv.org/abs/1210.7183}
  {arXiv:1210.7183 [astro-ph.CO]} \BibitemShut {NoStop}%
\bibitem [{\citenamefont {Metropolis}\ \emph {et~al.}(1953)\citenamefont
  {Metropolis}, \citenamefont {Rosenbluth}, \citenamefont {Rosenbluth},
  \citenamefont {Teller},\ and\ \citenamefont {Teller}}]{Metropolis:1953aaa}%
  \BibitemOpen
  \bibfield  {author} {\bibinfo {author} {\bibfnamefont {A.}~\bibnamefont
  {Metropolis}}, \bibinfo {author} {\bibfnamefont {A.}~\bibnamefont
  {Rosenbluth}}, \bibinfo {author} {\bibfnamefont {M.}~\bibnamefont
  {Rosenbluth}}, \bibinfo {author} {\bibfnamefont {A.}~\bibnamefont {Teller}},
  \ and\ \bibinfo {author} {\bibfnamefont {E.}~\bibnamefont {Teller}},\
  }\href@noop {} {\bibfield  {journal} {\bibinfo  {journal} {J. Chem. Phys.}\
  }\textbf {\bibinfo {volume} {21}},\ \bibinfo {pages} {1087} (\bibinfo {year}
  {1953})}\BibitemShut {NoStop}%
\bibitem [{\citenamefont {Hastings}(1970)}]{Hastings:1970bbb}%
  \BibitemOpen
  \bibfield  {author} {\bibinfo {author} {\bibfnamefont {W.}~\bibnamefont
  {Hastings}},\ }\href@noop {} {\bibfield  {journal} {\bibinfo  {journal}
  {Biometrika}\ }\textbf {\bibinfo {volume} {57}},\ \bibinfo {pages} {97}
  (\bibinfo {year} {1970})}\BibitemShut {NoStop}%
\bibitem [{\citenamefont {Gelman}\ and\ \citenamefont
  {Rubin}(1992)}]{GelmanRubin1992}%
  \BibitemOpen
  \bibfield  {author} {\bibinfo {author} {\bibfnamefont {A.}~\bibnamefont
  {Gelman}}\ and\ \bibinfo {author} {\bibfnamefont {D.~B.}\ \bibnamefont
  {Rubin}},\ }\href@noop {} {\bibfield  {journal} {\bibinfo  {journal}
  {Statistical Science}\ }\textbf {\bibinfo {volume} {7}},\ \bibinfo {pages}
  {457} (\bibinfo {year} {1992})}\BibitemShut {NoStop}%
\bibitem [{\citenamefont {S\'anchez}(2020)}]{Sanchez:2020vvb}%
  \BibitemOpen
  \bibfield  {author} {\bibinfo {author} {\bibfnamefont {A.~G.}\ \bibnamefont
  {S\'anchez}},\ }\href {\doibase 10.1103/PhysRevD.102.123511} {\bibfield
  {journal} {\bibinfo  {journal} {Phys. Rev. D}\ }\textbf {\bibinfo {volume}
  {102}},\ \bibinfo {pages} {123511} (\bibinfo {year} {2020})},\ \Eprint
  {http://arxiv.org/abs/2002.07829} {arXiv:2002.07829 [astro-ph.CO]}
  \BibitemShut {NoStop}%
\bibitem [{\citenamefont {G\'omez-Valent}\ \emph {et~al.}(2021)\citenamefont
  {G\'omez-Valent}, \citenamefont {Zheng}, \citenamefont {Amendola},
  \citenamefont {Pettorino},\ and\ \citenamefont
  {Wetterich}}]{Gomez-Valent:2021cbe}%
  \BibitemOpen
  \bibfield  {author} {\bibinfo {author} {\bibfnamefont {A.}~\bibnamefont
  {G\'omez-Valent}}, \bibinfo {author} {\bibfnamefont {Z.}~\bibnamefont
  {Zheng}}, \bibinfo {author} {\bibfnamefont {L.}~\bibnamefont {Amendola}},
  \bibinfo {author} {\bibfnamefont {V.}~\bibnamefont {Pettorino}}, \ and\
  \bibinfo {author} {\bibfnamefont {C.}~\bibnamefont {Wetterich}},\ }\href
  {\doibase 10.1103/PhysRevD.104.083536} {\bibfield  {journal} {\bibinfo
  {journal} {Phys. Rev. D}\ }\textbf {\bibinfo {volume} {104}},\ \bibinfo
  {pages} {083536} (\bibinfo {year} {2021})},\ \Eprint
  {http://arxiv.org/abs/2107.11065} {arXiv:2107.11065 [astro-ph.CO]}
  \BibitemShut {NoStop}%
\bibitem [{\citenamefont {Vagnozzi}(2021)}]{Vagnozzi:2021gjh}%
  \BibitemOpen
  \bibfield  {author} {\bibinfo {author} {\bibfnamefont {S.}~\bibnamefont
  {Vagnozzi}},\ }\href {\doibase 10.1103/PhysRevD.104.063524} {\bibfield
  {journal} {\bibinfo  {journal} {Phys. Rev. D}\ }\textbf {\bibinfo {volume}
  {104}},\ \bibinfo {pages} {063524} (\bibinfo {year} {2021})},\ \Eprint
  {http://arxiv.org/abs/2105.10425} {arXiv:2105.10425 [astro-ph.CO]}
  \BibitemShut {NoStop}%
\bibitem [{\citenamefont {Hill}\ \emph {et~al.}(2020)\citenamefont {Hill},
  \citenamefont {McDonough}, \citenamefont {Toomey},\ and\ \citenamefont
  {Alexander}}]{Hill:2020osr}%
  \BibitemOpen
  \bibfield  {author} {\bibinfo {author} {\bibfnamefont {J.~C.}\ \bibnamefont
  {Hill}}, \bibinfo {author} {\bibfnamefont {E.}~\bibnamefont {McDonough}},
  \bibinfo {author} {\bibfnamefont {M.~W.}\ \bibnamefont {Toomey}}, \ and\
  \bibinfo {author} {\bibfnamefont {S.}~\bibnamefont {Alexander}},\ }\href
  {\doibase 10.1103/PhysRevD.102.043507} {\bibfield  {journal} {\bibinfo
  {journal} {Phys. Rev. D}\ }\textbf {\bibinfo {volume} {102}},\ \bibinfo
  {pages} {043507} (\bibinfo {year} {2020})},\ \Eprint
  {http://arxiv.org/abs/2003.07355} {arXiv:2003.07355 [astro-ph.CO]}
  \BibitemShut {NoStop}%
\bibitem [{\citenamefont {D'Amico}\ \emph {et~al.}(2021)\citenamefont
  {D'Amico}, \citenamefont {Senatore}, \citenamefont {Zhang},\ and\
  \citenamefont {Zheng}}]{DAmico:2020ods}%
  \BibitemOpen
  \bibfield  {author} {\bibinfo {author} {\bibfnamefont {G.}~\bibnamefont
  {D'Amico}}, \bibinfo {author} {\bibfnamefont {L.}~\bibnamefont {Senatore}},
  \bibinfo {author} {\bibfnamefont {P.}~\bibnamefont {Zhang}}, \ and\ \bibinfo
  {author} {\bibfnamefont {H.}~\bibnamefont {Zheng}},\ }\href {\doibase
  10.1088/1475-7516/2021/05/072} {\bibfield  {journal} {\bibinfo  {journal}
  {JCAP}\ }\textbf {\bibinfo {volume} {05}},\ \bibinfo {pages} {072} (\bibinfo
  {year} {2021})},\ \Eprint {http://arxiv.org/abs/2006.12420} {arXiv:2006.12420
  [astro-ph.CO]} \BibitemShut {NoStop}%
\bibitem [{\citenamefont {Murgia}\ \emph {et~al.}(2021)\citenamefont {Murgia},
  \citenamefont {Abell\'an},\ and\ \citenamefont {Poulin}}]{Murgia:2020ryi}%
  \BibitemOpen
  \bibfield  {author} {\bibinfo {author} {\bibfnamefont {R.}~\bibnamefont
  {Murgia}}, \bibinfo {author} {\bibfnamefont {G.~F.}\ \bibnamefont
  {Abell\'an}}, \ and\ \bibinfo {author} {\bibfnamefont {V.}~\bibnamefont
  {Poulin}},\ }\href {\doibase 10.1103/PhysRevD.103.063502} {\bibfield
  {journal} {\bibinfo  {journal} {Phys. Rev. D}\ }\textbf {\bibinfo {volume}
  {103}},\ \bibinfo {pages} {063502} (\bibinfo {year} {2021})},\ \Eprint
  {http://arxiv.org/abs/2009.10733} {arXiv:2009.10733 [astro-ph.CO]}
  \BibitemShut {NoStop}%
\bibitem [{\citenamefont {Hill}\ \emph {et~al.}(2021)\citenamefont {Hill} \emph
  {et~al.}}]{Hill:2021yec}%
  \BibitemOpen
  \bibfield  {author} {\bibinfo {author} {\bibfnamefont {J.~C.}\ \bibnamefont
  {Hill}} \emph {et~al.},\ }\href@noop {} {\  (\bibinfo {year} {2021})},\
  \Eprint {http://arxiv.org/abs/2109.04451} {arXiv:2109.04451 [astro-ph.CO]}
  \BibitemShut {NoStop}%
\bibitem [{\citenamefont {Poulin}\ \emph {et~al.}(2021)\citenamefont {Poulin},
  \citenamefont {Smith},\ and\ \citenamefont {Bartlett}}]{Poulin:2021bjr}%
  \BibitemOpen
  \bibfield  {author} {\bibinfo {author} {\bibfnamefont {V.}~\bibnamefont
  {Poulin}}, \bibinfo {author} {\bibfnamefont {T.~L.}\ \bibnamefont {Smith}}, \
  and\ \bibinfo {author} {\bibfnamefont {A.}~\bibnamefont {Bartlett}},\ }\href
  {\doibase 10.1103/PhysRevD.104.123550} {\bibfield  {journal} {\bibinfo
  {journal} {Phys. Rev. D}\ }\textbf {\bibinfo {volume} {104}},\ \bibinfo
  {pages} {123550} (\bibinfo {year} {2021})},\ \Eprint
  {http://arxiv.org/abs/2109.06229} {arXiv:2109.06229 [astro-ph.CO]}
  \BibitemShut {NoStop}%
\bibitem [{\citenamefont {Aiola}\ \emph {et~al.}(2020)\citenamefont {Aiola}
  \emph {et~al.}}]{ACT:2020gnv}%
  \BibitemOpen
  \bibfield  {author} {\bibinfo {author} {\bibfnamefont {S.}~\bibnamefont
  {Aiola}} \emph {et~al.} (\bibinfo {collaboration} {ACT}),\ }\href {\doibase
  10.1088/1475-7516/2020/12/047} {\bibfield  {journal} {\bibinfo  {journal}
  {JCAP}\ }\textbf {\bibinfo {volume} {12}},\ \bibinfo {pages} {047} (\bibinfo
  {year} {2020})},\ \Eprint {http://arxiv.org/abs/2007.07288} {arXiv:2007.07288
  [astro-ph.CO]} \BibitemShut {NoStop}%
\bibitem [{\citenamefont {G\'omez-Valent}(2022)}]{Gomez-Valent:2021hda}%
  \BibitemOpen
  \bibfield  {author} {\bibinfo {author} {\bibfnamefont {A.}~\bibnamefont
  {G\'omez-Valent}},\ }\href {\doibase 10.1103/PhysRevD.105.043528} {\bibfield
  {journal} {\bibinfo  {journal} {Phys. Rev. D}\ }\textbf {\bibinfo {volume}
  {105}},\ \bibinfo {pages} {043528} (\bibinfo {year} {2022})},\ \Eprint
  {http://arxiv.org/abs/2111.15450} {arXiv:2111.15450 [astro-ph.CO]}
  \BibitemShut {NoStop}%
\bibitem [{\citenamefont {Freedman}(2021)}]{Freedman:2021ahq}%
  \BibitemOpen
  \bibfield  {author} {\bibinfo {author} {\bibfnamefont {W.~L.}\ \bibnamefont
  {Freedman}},\ }\href {\doibase 10.3847/1538-4357/ac0e95} {\bibfield
  {journal} {\bibinfo  {journal} {Astrophys. J.}\ }\textbf {\bibinfo {volume}
  {919}},\ \bibinfo {pages} {16} (\bibinfo {year} {2021})},\ \Eprint
  {http://arxiv.org/abs/2106.15656} {arXiv:2106.15656 [astro-ph.CO]}
  \BibitemShut {NoStop}%
\bibitem [{\citenamefont {Joudaki}\ \emph {et~al.}(2020)\citenamefont {Joudaki}
  \emph {et~al.}}]{Joudaki:2019pmv}%
  \BibitemOpen
  \bibfield  {author} {\bibinfo {author} {\bibfnamefont {S.}~\bibnamefont
  {Joudaki}} \emph {et~al.},\ }\href {\doibase 10.1051/0004-6361/201936154}
  {\bibfield  {journal} {\bibinfo  {journal} {Astron. Astrophys.}\ }\textbf
  {\bibinfo {volume} {638}},\ \bibinfo {pages} {L1} (\bibinfo {year} {2020})},\
  \Eprint {http://arxiv.org/abs/1906.09262} {arXiv:1906.09262 [astro-ph.CO]}
  \BibitemShut {NoStop}%
\bibitem [{\citenamefont {Chang}\ \emph {et~al.}(2022)\citenamefont {Chang}
  \emph {et~al.}}]{DES:2022ign}%
  \BibitemOpen
  \bibfield  {author} {\bibinfo {author} {\bibfnamefont {C.}~\bibnamefont
  {Chang}} \emph {et~al.} (\bibinfo {collaboration} {DES, SPT}),\ }\href@noop
  {} {\  (\bibinfo {year} {2022})},\ \Eprint {http://arxiv.org/abs/2203.12440}
  {arXiv:2203.12440 [astro-ph.CO]} \BibitemShut {NoStop}%
\bibitem [{\citenamefont {Moss}\ \emph {et~al.}(2021)\citenamefont {Moss},
  \citenamefont {Copeland}, \citenamefont {Bamford},\ and\ \citenamefont
  {Clarke}}]{Moss:2021obd}%
  \BibitemOpen
  \bibfield  {author} {\bibinfo {author} {\bibfnamefont {A.}~\bibnamefont
  {Moss}}, \bibinfo {author} {\bibfnamefont {E.}~\bibnamefont {Copeland}},
  \bibinfo {author} {\bibfnamefont {S.}~\bibnamefont {Bamford}}, \ and\
  \bibinfo {author} {\bibfnamefont {T.}~\bibnamefont {Clarke}},\ }\href@noop {}
  {\  (\bibinfo {year} {2021})},\ \Eprint {http://arxiv.org/abs/2109.14848}
  {arXiv:2109.14848 [astro-ph.CO]} \BibitemShut {NoStop}%
\bibitem [{\citenamefont {Fondi}\ \emph {et~al.}(2022)\citenamefont {Fondi},
  \citenamefont {Melchiorri},\ and\ \citenamefont {Pagano}}]{Fondi:2022tfp}%
  \BibitemOpen
  \bibfield  {author} {\bibinfo {author} {\bibfnamefont {E.}~\bibnamefont
  {Fondi}}, \bibinfo {author} {\bibfnamefont {A.}~\bibnamefont {Melchiorri}}, \
  and\ \bibinfo {author} {\bibfnamefont {L.}~\bibnamefont {Pagano}},\
  }\href@noop {} {\bibfield  {journal} {\bibinfo  {journal} {Astrophys. J.
  Lett.}\ }\textbf {\bibinfo {volume} {931}},\ \bibinfo {pages} {L18} (\bibinfo
  {year} {2022})},\ \Eprint {http://arxiv.org/abs/2203.12930} {arXiv:2203.12930
  [astro-ph.CO]} \BibitemShut {NoStop}%
\bibitem [{\citenamefont {Xia}(2013)}]{Xia:2013nua}%
  \BibitemOpen
  \bibfield  {author} {\bibinfo {author} {\bibfnamefont {J.-Q.}\ \bibnamefont
  {Xia}},\ }\href {\doibase 10.1088/1475-7516/2013/11/022} {\bibfield
  {journal} {\bibinfo  {journal} {JCAP}\ }\textbf {\bibinfo {volume} {1311}},\
  \bibinfo {pages} {022} (\bibinfo {year} {2013})},\ \Eprint
  {http://arxiv.org/abs/1311.2131} {arXiv:1311.2131 [astro-ph.CO]} \BibitemShut
  {NoStop}%
\bibitem [{\citenamefont {van~de Bruck}\ \emph {et~al.}(2017)\citenamefont
  {van~de Bruck}, \citenamefont {Mifsud},\ and\ \citenamefont
  {Morrice}}]{vandeBruck:2016hpz}%
  \BibitemOpen
  \bibfield  {author} {\bibinfo {author} {\bibfnamefont {C.}~\bibnamefont
  {van~de Bruck}}, \bibinfo {author} {\bibfnamefont {J.}~\bibnamefont
  {Mifsud}}, \ and\ \bibinfo {author} {\bibfnamefont {J.}~\bibnamefont
  {Morrice}},\ }\href {\doibase 10.1103/PhysRevD.95.043513} {\bibfield
  {journal} {\bibinfo  {journal} {Phys. Rev.}\ }\textbf {\bibinfo {volume}
  {D95}},\ \bibinfo {pages} {043513} (\bibinfo {year} {2017})},\ \Eprint
  {http://arxiv.org/abs/1609.09855} {arXiv:1609.09855 [astro-ph.CO]}
  \BibitemShut {NoStop}%
\bibitem [{\citenamefont {van~de Bruck}\ and\ \citenamefont
  {Mifsud}(2018)}]{vandeBruck:2017idm}%
  \BibitemOpen
  \bibfield  {author} {\bibinfo {author} {\bibfnamefont {C.}~\bibnamefont
  {van~de Bruck}}\ and\ \bibinfo {author} {\bibfnamefont {J.}~\bibnamefont
  {Mifsud}},\ }\href {\doibase 10.1103/PhysRevD.97.023506} {\bibfield
  {journal} {\bibinfo  {journal} {Phys. Rev.}\ }\textbf {\bibinfo {volume}
  {D97}},\ \bibinfo {pages} {023506} (\bibinfo {year} {2018})},\ \Eprint
  {http://arxiv.org/abs/1709.04882} {arXiv:1709.04882 [astro-ph.CO]}
  \BibitemShut {NoStop}%
\bibitem [{\citenamefont {Agrawal}\ \emph {et~al.}(2021)\citenamefont
  {Agrawal}, \citenamefont {Obied},\ and\ \citenamefont
  {Vafa}}]{Agrawal:2019dlm}%
  \BibitemOpen
  \bibfield  {author} {\bibinfo {author} {\bibfnamefont {P.}~\bibnamefont
  {Agrawal}}, \bibinfo {author} {\bibfnamefont {G.}~\bibnamefont {Obied}}, \
  and\ \bibinfo {author} {\bibfnamefont {C.}~\bibnamefont {Vafa}},\ }\href
  {\doibase 10.1103/PhysRevD.103.043523} {\bibfield  {journal} {\bibinfo
  {journal} {Phys. Rev. D}\ }\textbf {\bibinfo {volume} {103}},\ \bibinfo
  {pages} {043523} (\bibinfo {year} {2021})},\ \Eprint
  {http://arxiv.org/abs/1906.08261} {arXiv:1906.08261 [astro-ph.CO]}
  \BibitemShut {NoStop}%
\bibitem [{\citenamefont {Reid}\ \emph {et~al.}(2019)\citenamefont {Reid},
  \citenamefont {Pesce},\ and\ \citenamefont {Riess}}]{Reid:2019tiq}%
  \BibitemOpen
  \bibfield  {author} {\bibinfo {author} {\bibfnamefont {M.~J.}\ \bibnamefont
  {Reid}}, \bibinfo {author} {\bibfnamefont {D.~W.}\ \bibnamefont {Pesce}}, \
  and\ \bibinfo {author} {\bibfnamefont {A.~G.}\ \bibnamefont {Riess}},\ }\href
  {\doibase 10.3847/2041-8213/ab552d} {\bibfield  {journal} {\bibinfo
  {journal} {Astrophys. J. Lett.}\ }\textbf {\bibinfo {volume} {886}},\
  \bibinfo {pages} {L27} (\bibinfo {year} {2019})},\ \Eprint
  {http://arxiv.org/abs/1908.05625} {arXiv:1908.05625 [astro-ph.GA]}
  \BibitemShut {NoStop}%
\bibitem [{\citenamefont {Bertotti}\ \emph {et~al.}(2003)\citenamefont
  {Bertotti}, \citenamefont {Iess},\ and\ \citenamefont
  {Tortora}}]{Bertotti:2003rm}%
  \BibitemOpen
  \bibfield  {author} {\bibinfo {author} {\bibfnamefont {B.}~\bibnamefont
  {Bertotti}}, \bibinfo {author} {\bibfnamefont {L.}~\bibnamefont {Iess}}, \
  and\ \bibinfo {author} {\bibfnamefont {P.}~\bibnamefont {Tortora}},\ }\href
  {\doibase 10.1038/nature01997} {\bibfield  {journal} {\bibinfo  {journal}
  {Nature}\ }\textbf {\bibinfo {volume} {425}},\ \bibinfo {pages} {374}
  (\bibinfo {year} {2003})}\BibitemShut {NoStop}%
\bibitem [{\citenamefont {G\'omez-Valent}\ and\ \citenamefont
  {Hassan~Puttasiddappa}(2021)}]{Gomez-Valent:2021joz}%
  \BibitemOpen
  \bibfield  {author} {\bibinfo {author} {\bibfnamefont {A.}~\bibnamefont
  {G\'omez-Valent}}\ and\ \bibinfo {author} {\bibfnamefont {P.}~\bibnamefont
  {Hassan~Puttasiddappa}},\ }\href {\doibase 10.1088/1475-7516/2021/09/040}
  {\bibfield  {journal} {\bibinfo  {journal} {JCAP}\ }\textbf {\bibinfo
  {volume} {09}},\ \bibinfo {pages} {040} (\bibinfo {year} {2021})},\ \Eprint
  {http://arxiv.org/abs/2105.14819} {arXiv:2105.14819 [astro-ph.CO]}
  \BibitemShut {NoStop}%
\bibitem [{\citenamefont {Marra}\ and\ \citenamefont
  {Perivolaropoulos}(2021)}]{Marra:2021fvf}%
  \BibitemOpen
  \bibfield  {author} {\bibinfo {author} {\bibfnamefont {V.}~\bibnamefont
  {Marra}}\ and\ \bibinfo {author} {\bibfnamefont {L.}~\bibnamefont
  {Perivolaropoulos}},\ }\href {\doibase 10.1103/PhysRevD.104.L021303}
  {\bibfield  {journal} {\bibinfo  {journal} {Phys. Rev. D}\ }\textbf {\bibinfo
  {volume} {104}},\ \bibinfo {pages} {L021303} (\bibinfo {year} {2021})},\
  \Eprint {http://arxiv.org/abs/2102.06012} {arXiv:2102.06012 [astro-ph.CO]}
  \BibitemShut {NoStop}%
\bibitem [{\citenamefont {Amendola}\ and\ \citenamefont
  {G\'omez-Valent}(2020)}]{Amendola:2020qkb}%
  \BibitemOpen
  \bibfield  {author} {\bibinfo {author} {\bibfnamefont {L.}~\bibnamefont
  {Amendola}}\ and\ \bibinfo {author} {\bibfnamefont {A.}~\bibnamefont
  {G\'omez-Valent}},\ }\href {\doibase 10.1093/mnras/staa2362} {\bibfield
  {journal} {\bibinfo  {journal} {Mon. Not. Roy. Astron. Soc.}\ }\textbf
  {\bibinfo {volume} {498}},\ \bibinfo {pages} {181} (\bibinfo {year}
  {2020})},\ \Eprint {http://arxiv.org/abs/2007.02615} {arXiv:2007.02615
  [astro-ph.CO]} \BibitemShut {NoStop}%
\bibitem [{\citenamefont {Rizzato}\ and\ \citenamefont
  {Sellentin}(2022)}]{Rizzato:2022hbu}%
  \BibitemOpen
  \bibfield  {author} {\bibinfo {author} {\bibfnamefont {M.}~\bibnamefont
  {Rizzato}}\ and\ \bibinfo {author} {\bibfnamefont {E.}~\bibnamefont
  {Sellentin}},\ }\href@noop {} {\  (\bibinfo {year} {2022})},\ \Eprint
  {http://arxiv.org/abs/2203.05009} {arXiv:2203.05009 [astro-ph.CO]}
  \BibitemShut {NoStop}%
\bibitem [{\citenamefont {Powell}(2009)}]{BOBYQA}%
  \BibitemOpen
  \bibfield  {author} {\bibinfo {author} {\bibfnamefont {M.~J.~D.}\
  \bibnamefont {Powell}},\ }\href@noop {} {\  (\bibinfo {year} {2009})},\
  \bibinfo {note} {technical report (DAMTP 2009/NA06), Centre for Mathematical
  Sciences, University of Cambridge, UK}\BibitemShut {NoStop}%
\bibitem [{\citenamefont {Cartis}\ \emph {et~al.}(2018)\citenamefont {Cartis},
  \citenamefont {Fiala}, \citenamefont {Marteau},\ and\ \citenamefont
  {Roberts}}]{PyBOBYQA}%
  \BibitemOpen
  \bibfield  {author} {\bibinfo {author} {\bibfnamefont {C.}~\bibnamefont
  {Cartis}}, \bibinfo {author} {\bibfnamefont {J.}~\bibnamefont {Fiala}},
  \bibinfo {author} {\bibfnamefont {B.}~\bibnamefont {Marteau}}, \ and\
  \bibinfo {author} {\bibfnamefont {L.}~\bibnamefont {Roberts}},\ }\href@noop
  {} {\  (\bibinfo {year} {2018})},\ \bibinfo {note} {technical report,
  University of Oxford, UK}\BibitemShut {NoStop}%
\end{thebibliography}%

\end{document}